\documentclass[useAMS,usenatbib]{mn2e}
\usepackage{graphicx}

\title[Foreground removal using {\sc fastica}]{Foreground Removal using {\sc fastica}: A Showcase of LOFAR-EoR}
\author[E. Chapman et al.]
{Emma Chapman,$^1$\thanks{eow@star.ucl.ac.uk}
 Filipe B. Abdalla,$^1$ Geraint Harker,$^{2,3}$ Vibor Jeli\'{c},$^{4}$
\newauthor
Panagiotis Labropoulos,$^{4,5}$ Saleem Zaroubi,$^5$ Michiel A. Brentjens, $^4$ 
\newauthor 
A.G. de Bruyn, $^{4,5}$ L.V.E.Koopmans,$^5$  \\ 
$^1$Department of Physics \& Astronomy, University College London, Gower Street, London, WC1E 6BT \\
$^2$Center for Astrophysics and Space Astronomy, 389 UCB, University of Colorado, Boulder, CO 80309-0389, USA \\
$^3$NASA Lunar Science Institute, NASA Ames Research Center, Moffett Field, CA 94035, USA \\
$^4$ASTRON, PO Box 2, NL-7990AA Dwingeloo, the Netherlands\\
$^5$Kapteyn Astronomical Institute, University of Groningen, PO Box 800, 9700AV Groningen, the Netherlands }

\def\LaTeX{L\kern-.36em\raise.3ex\hbox{a}\kern-.15em
    T\kern-.1667em\lower.7ex\hbox{E}\kern-.125emX}

\begin{document}

\maketitle

\begin{abstract}
We introduce a new implementation of the \textsc{fastica} algorithm on simulated LOFAR-EoR data with the aim of accurately removing the foregrounds and extracting the 21-cm reionization signal. We find that the method successfully removes the foregrounds with an average fitting error of 0.5 per cent and that the 2D and 3D power spectra are recovered across the frequency range. We find that for scales above several PSF scales the 21-cm variance is successfully recovered though there is evidence of noise leakage into the reconstructed foreground components. We find that this blind independent component analysis technique provides encouraging results without the danger of prior foreground assumptions. 
\end{abstract}


\section{Introduction}
Four hundred thousand years after the Big Bang, the recombination of electrons and protons resulted in a neutral Universe, steadily cooling with the Hubble expansion. The `Dark Ages' followed recombination until, 400 million years after the Big Bang, the first ionizing sources came into existence. This Epoch of Reionization (EoR) is one of the last unobserved eras of our Universe, but with a new generation of radio telescopes coming online (e.g. Low Frequency Array (LOFAR)\footnote{http://www.lofar.org/}, Giant Metrewave Radio Telescope (GMRT)\footnote{http://gmrt.ncra.tifr.res.in/}, Murchison Widefield Array (MWA)\footnote{http://www.mwatelescope.org/},Precision Array to Probe the Epoch of Reionization (PAPER)\footnote{http://astro.berkeley.edu/~dbacker/eor/}, 21 Centimeter Array (21CMA)\footnote{http://21cma.bao.ac.cn/}, Square Kilometre Array (SKA)\footnote{http://www.skatelescope.org/}) this is soon about to change.

It is generally accepted that the most rewarding way to probe reionization is through the 21-cm spectral line - produced by a spin flip in neutral hydrogen \citep{vandehulst45,ewen51,muller51}. This 21-cm radiation can be observed interferometrically at radio wavelengths as a deviation from the brightness temperature of the CMB (\citealt{field58}; \citealt{field59}; \citealt*{madau97}; \citealt{shaver99}). 

Observationally, the 21-cm signal will be accompanied by system noise and Galactic and extragalactic foregrounds \citep[e.g.][]{jelic08,jelic10} which are orders of magnitude larger than the 21-cm signal we wish to detect. On top of this there are systematic effects due to the ionosphere and instrument response. The foregrounds must be carefully removed using a cleaning process of high accuracy and precision as any error at this stage has the ability to destroy the underlying 21-cm signal. Foreground removal and the implications for 21-cm cosmology has been extensively researched over the past decade (e.g \citealt{dimatteo02}; \citealt{oh03}; \citealt*{dimatteo04}; \citealt*{zal04}; \citealt{morales04}; \citealt*{santos05}; \citealt{wang06}; \citealt{mcquinn06}; \citealt{jelic08}; \citealt*{gleser08}; \citealt*{bowman06}; \citealt{harker09b}; \citealt*{liu09a}; \citealt{liu09b}; \citealt{harker10}; \citealt{liu11}; \citealt{petrovic11}; \citealt{mao12}; \citealt{liu12}). This paper constitutes only one step in the foreground removal process and assumes that bright sources have been removed, for example via a flux cut \citep{dimatteo04}. 

There is currently no consensus on the most effective foreground removal method, though a recent method implemented by \citet{harker09b,harker10} has shown promising results while making only minimal assumptions. The same method highlighted that foreground removal techniques can be carried out both in image and visibility space - with the quality of results sometimes dependent on the choice of space. It is possible that different methods will be best suited for the extraction of different information from the data and improvements in the recovery of any one statistic, even at the expense of another, will still be useful. The statistical detection of the EoR signal is fraught with uncertainty and applying several foreground cleaning methods to the data independently will be invaluable in confirming a statistical detection. 

The method presented here is based on the independent component analysis (ICA) algorithm, \textsc{fastica} \citep*{hyvarinen01}. ICA is a method originally designed to separate independent signals with minimal prior knowledge of the form of the signals. Thus ICA provides us with a foreground removal method which compensates for the fact that we do not know the form of the foregrounds at the exact resolution and frequency range of LOFAR. A non-parametric method, \textsc{fastica} allows the foregrounds to choose their own shape instead of assuming a specific form, such as a polynomial. \textsc{fastica} is a versatile tool and has been applied recently in the field of exoplanets \citep{waldmann12} and CMB foreground removal with great success (e.g. \citealt{maino02}; \citealt{maino03}; \citealt{maino07}; \citealt*{bottino08}; \citealt*{bottino10}) motivating its implementation on other cosmological data. The results presented focus on the two main statistical aims of current EoR experiments, namely the recovery of the power spectrum and variance of the cosmological signal.

Section \ref{fastica} briefly describes the \textsc{fastica} methodology and algorithm used to identify independent components (ICs). The various methods used to simulate the 21-cm signal, foregrounds and noise are set out in Section \ref{sims}. The results and sensitivity of the \textsc{fastica} method are presented in Sections \ref{results} and \ref{limit}  before a final summary and discussion in Section \ref{conclusions}.

\section{Foreground Removal Techniques}
\label{fastica}
The statistical detection of the 21-cm reionization signal depends on an accurate and robust method for removing the foregrounds from the total signal. Since it is impossible to observe the foregrounds alone this is not a simple task.
 
The first attempts focused on exploiting the angular fluctuations of the 21-cm signal \citep[e.g.][]{dimatteo02,oh03,dimatteo04}, but the 21-cm signal was found to be swamped by various foregrounds. The focus then moved on to the frequency correlation of the foregrounds, with the cross-correlation of pairs of maps used as a cleaning step \citep{zal04,santos05}. While the foregrounds are expected to be highly correlated on scales of 1 MHz \citep[e.g.][]{gnedin04,dimatteo02}, the cosmological signal is expected to be highly uncorrelated \citep*{ali08} on the same frequency scales, allowing frequency correlation to differentiate the signals.

As more methods have emerged, it has become clear that different methods have different advantages and foreground subtraction has become accepted as a three stage process. The first and last stages are bright source removal (\citealt{cooray04}; \citealt{dimatteo04}; \citealt{zal04}; \citealt*{morales06}) and residual error subtraction \citep{morales04} and are not dealt with in this paper. For our data we assume that the first stage has been carried out and all bright sources have been removed, which will still leave foregrounds strong enough to swamp the 21-cm signal \citep{dimatteo02,oh03}. The second stage is known as spectral fitting, or line of sight fitting, and has become by far the most popular in literature. Line of sight methods can be divided into subcategories of parametric and non-parametric methods. The majority of literature involves parametric methods, whereby at some point a certain form for the foregrounds is assumed, for example polynomials (e.g. \citealt{santos05}; \citealt{wang06}; \citealt{mcquinn06}; \citealt{bowman06}; \citealt{jelic08}; \citealt{gleser08}; \citealt*{liu09a}; \citealt{liu09b}; \citealt{petrovic11}). In contrast, non-parametric methods allow the data to determine the form of the foregrounds with many more free parameters, allowing much more freedom and not assuming a specific form. This has obvious advantages for a cosmological era so far not directly observed but results are often not as promising as parametric results. A possible exception is the recent method presented by \citet{harker09b,harker10} which preferentially considered foreground models with as few inflection points as possible. When applied to LOFAR-EoR data, this method compared very favourably with parametric methods. \textsc{fastica} is another non-parametric method which we will show produces similarly promising results.

\subsection{The {\sc fastica} method}
\subsubsection{Background}
Introduced in the early 1980s, ICA has established itself as a successful component separation technique with widespread applications. The method relies on the assumption that the multiple elements making up a mixed signal are statistically independent.

ICA methods often formulate the data model as:

\begin{equation}
\label{xas}
\bmath{x}=\mathbfss{A}\bmath{s}
\end{equation}

\noindent where $\bmath{x}$ is a vector representing the observed signal, $\bmath{s}$ a vector of which the components are assumed mutually independent and $\mathbfss{A}$ a mixing matrix to be calculated. For our data we have signal maps of $512\times512$ pixels at 170 different frequencies. Equation \ref{xas} represents one line of sight where, if $m$ ICs are assumed, the sizes of $\bmath{x}$, $\mathbfss{A}$ and $\bmath{s}$ are [170,1], [170,$m$] and [$m$,1] respectively. Actually, \textsc{fastica} simultaneously considers all lines of sight, so $\bmath{x}$ and $\bmath{s}$ are in effect matrices of size [170,512 $\times$ 512] and [$m$,512 $\times$ 512] respectively. For clarity, we will set out the description as if only one line of sight was being considered but the reader should bear in mind that all lines of sight are simultaneously and independently treated by the algorithm, with $\mathbfss{A}$ being independent of the line of sight.

It might immediately strike the reader that the model specified here is the noise free ICA model. This is because this implementation makes no effort to model the noise component through the $\bmath{x}=\mathbfss{A}\bmath{s} + \bmath{n}$ formulation. Instead one must appreciate that it is the way in which \textsc{fastica} is not robust to noisy components that we take advantage of here. Whereas $\bmath{x}$ will represent the observed signal of foreground, noise and 21-cm signal, $\bmath{s}$ is considered to be the foregrounds only. \textsc{fastica} ignores the Gaussian or non-smooth spectral components in the observed signal. When we specify $m$ ICs, \textsc{fastica} reconstructs $m$ ICs relating to the foregrounds only. 

To solve Equation \ref{xas} for $\bmath{s}$, we seek a linear transform:

\begin{equation}
\bmath{s}=\mathbfss{W}\bmath{x}
\end{equation}

\noindent where $\mathbfss{W}$ is a constant weight matrix that the ICA method aims to determine assuming the elements of $\bmath{s}$ are as statistically independent as possible.

\textsc{fastica} seeks to estimate $\mathbfss{W}$ using the concept of mutual information. An outline of the general philosophy behind \textsc{fastica} is outlined below, but for a full treatment refer to \citet{hyvarinen99} and \citet{hyvarinen01}.  

Let us consider a single component of the signal $\bmath{s}$:

\begin{equation}
\label{y}
y=\bmath{w}^T\bmath{x}=\sum_i{w_i x_i}
\end{equation}

\noindent where if $\bmath{w}$ is one of the rows of the inverse of $\mathbfss{A}$, $y$ is actually one of the ICs, $s_i$. ICA then attempts to minimise the Gaussianity of $\bmath{w}^T\bmath{x}$. To understand why, we define a vector $\bmath{z}$ :

\begin{equation}
\bmath{z}=\mathbfss{A}^T\bmath{w}
\end{equation}

\noindent so that we have a weighted sum of the independent signal components:

\begin{equation}
y=\bmath{z}^T\bmath{s}
\end{equation}

\noindent The central limit theorem states that the greater the number of independent variables in a distribution, the more Gaussian that distribution will be. $\bmath{z}^T\bmath{s}$ is therefore always more Gaussian than any individual $s_i$. $y$ will be least Gaussian when one, and only one, $z_i$ is non-zero, and in such a case $y$ is then one of the ICs. Thus by maximising the non-Gaussianity of $\bmath{w}^T\bmath{x}$ we find one of the ICs.

In order to estimate and adjust $\bmath{w}^T\bmath{x}$ in such a way that its Gaussianity converges to a minimum, the methods needs a robust measure of non-Gaussianity. \textsc{fastica} favours negentropy as a measure of non-Gaussianity, which is based on the idea of the entropy of a variable, $H(y)$:

\begin{equation}
H(y) = - \sum_i P(y=a_i) \log P (y=a_i)
\end{equation}

\noindent where $a_i$ are the possible values of $y$.

Negentropy is then defined as:

\begin{equation}
J(y)=H(y_{\mathrm{gauss}}) - H(y)
\end{equation}

\noindent where $y_{\mathrm{gauss}}$ is a random Gaussian variable with the same covariance matrix as $y$. Using the maximum-entropy principle, one can define:

\begin{equation}
J(y) \approx \sum_{i=1}^n k_i [E\{G_i(y)\}-E\{G_i(\nu)\}]^2   
\end{equation}

\noindent where $k_i$ are positive constants, $\nu$ is a Gaussian variable with zero mean and unit variance and $G$ are non-quadratic functions. Though almost any non-quadratic function can be used, the robustness and speed of the \textsc{fastica} method depends on choosing these contrast functions well, with different contrast functions more suited to different scenarios. For this implementation we choose a non linearity, $g(u)$ of:

\begin{equation} 
g(u) = u \times \exp\left(-\frac{u^2}{2}\right)
\end{equation} 
\noindent where $g(u)=G'(u)=\frac{dG(u)}{du}$. This choice is particularly suited when robustness is important or when the components have high kurtosis.

Since $\bmath{s}$ and $\mathbfss{A}$ are both unknown, \textsc{fastica} cannot determine the ICs' magnitudes or order, as we can freely change the order of the components in the mixing model or multiply any of them by a scalar factor which can be balanced out by dividing out elsewhere. As such \textsc{fastica} fixes the magnitudes of the ICs by assuming they have unit variance.

\subsubsection{Algorithm}
Here we summarise the fixed-point \textsc{fastica} algorithm for finding one IC. 

The mixed signal is input along with a parameter representing the number of ICs we assume there to be. A typical choice in this implementation is four ICs.

This data undergoes several preprocessing steps within the \textsc{fastica} program. First the data are adjusted to be of zero mean to simplify the algorithm. Then, using a principal component analysis to estimate the eigenvalues and eigenvectors for the data, the data are whitened. This results in the vector $\bmath{x}$ where the components are uncorrelated, with unit variances.

We wish to choose a unit vector $\bmath{w}$ such that the non-Gaussianity of $\bmath{w}^T\bmath{x}$ is maximized. Under the assumption that the components have unit variance (which for whitened data is equivalent to assuming $||\bmath{w}||^2$=1) these maxima occur where:

\begin{equation}
E\{\bmath{x}g(\bmath{w}^T\bmath{x})\}-\beta \bmath{w} = 0
\end{equation}

\noindent is satisfied. To find the roots of this equation using Newton's method we arrive at the approximate Newton iteration:

\begin{equation}
\bmath{w}^+ = \bmath{w} - \frac{[E\{\bmath{x}g(\bmath{w}^T\bmath{x})\}-\beta \bmath{w}]}{E\{g'(\bmath{w}^T\bmath{x})\}-\beta}
\end{equation}

This iteration is carried out using the algorithm summarised in the following steps \citep{hyvarinen00}:

\begin{enumerate}
\item Choose an initial random weight vector $\bmath{w}$
\item Let $\bmath{w}^+ = E\{\bmath{x}g(\bmath{w}^T\bmath{x})\}- E\{g'(\bmath{w}^T\bmath{x})\}\bmath{w}$
\item Let $\bmath{w}=\frac{\bmath{w}^+}{\parallel\bmath{w}^+\parallel}$
\item If the old and new values for $\bmath{w}$ are not converged repeat the process
\end{enumerate}

\noindent where $g$ is the derivative and $g'$ is the second derivative of the chosen contrast function $G$. The use of the contrast function derivatives comes from consideration of where the maxima of the negentropy approximation are obtained. The non-Gaussianity is maximised along the line of sight and across the map simultaneously meaning that the method's constraining power benefits from having more pixels and more frequency maps. 

To extend the algorithm to $n$ components requires \textsc{fastica} to run simultaneously for $n$ different weight vectors, $\bmath{w}_1,...\bmath{w}_n$ where one $\bmath{w}_j$ corresponds to $\bmath{w}$ in the above algorithm. To ensure that the different $\bmath{w_j} \bmath{x}$ converge to different maxima (i.e. the same IC is not found twice) all the outputs $\bmath{w_j} \bmath{x}$ must be decorrelated after every iteration. For a more detailed treatment refer to \citet{hyvarinen99} and \citet{hyvarinen01}.

\subsubsection{Our Implementation}
We make use of the C++ implementation of \textsc{fastica} provided by the IT++ library\footnote{http://itpp.sourceforge.net/devel/fastica\_8cpp.html}. Our foreground subtraction proceeds in the following steps:

\begin{enumerate}
\item Read in the simulation data cube and specify the number of foreground ICs for \textsc{fastica} to model.
\item Call \textsc{fastica} to calculate the mixing matrix and ICs of the foregrounds. 
\item Reconstruct the foregrounds by performing a multiplication of the mixing matrix, which is common to all  lines of sight, with the vector of ICs for each line of sight.
\item Find the difference between the reconstructed foreground cube and the input cube. This residual cube should equal the 21-cm signal, noise and and any foreground fitting errors.
\end{enumerate}

Statistical tests can then be carried out on the residuals cube to determine if the 21-cm signal is recoverable after the foreground removal process.

It is worth reiterating once more as it such an important point: the ICs referred to in the ICA methodology as applied here are the ICs making up the foregrounds - the cosmological signal and noise are at no point modelled or even taken into account by this \textsc{fastica} implementation. 
\section{Simulated EoR Data}
We simulate 170 frequency maps between 115 and 200 MHz with spacings of 0.5 MHz. The maps consisted of $512^2$ pixels representing a 10$^{\circ} \times$ 10$^{\circ}$ observation window, or a resolution of 1.17 arcminutes per pixel. Since an interferometer like LOFAR is insensitive to the mean value of the brightness temperature, we use mean-subtracted maps.
\label{sims}

\subsection{21-cm Cosmological Signal}
Of the existing reionization simulation programs \citep[e.g.][]{santos10}, we use the semi-numeric modelling tool 21cmFAST to simulate the observable of the 21-cm radiation, the brightness temperature $T_\mathrm{b}$ (\citealt{mesinger07}; \citealt*{mesinger11}). 21cmFAST treats physical processes with approximate methods for small realisation generation times and has produced results which agree well with the most recent hydrodynamical simulations. The code was run using a standard cosmology, ($\Omega_{\Lambda},\Omega_m,\Omega_b,n,\sigma_8,h$)=(0.72,0.28,0.046,0.96,0.82,73)\citep{komatsu11} and initialised at $z$=300 on a $1800^3$ grid. The velocity fields used to perturb the initial conditions as well as the resulting 21-cm $T_\mathrm{b}$ boxes were formed on a cruder grid of $450^3$ before being interpolated up to $512^3$. We define halos contributing ionizing photons as having a minimum virial mass of $1\times 10^9 \mathrm{M\odot}$.

When a hydrogen atom undergoes a ground state hyperfine transition from the excited triplet state, where the spins of the proton and electron are parallel, to the singlet state, where the spins are anti-parallel, a photon is emitted of wavelength 21-cm or frequency $\nu_{10}=1420\:\mathrm{MHz}$. The 21-cm spectral line is forbidden - the probability of a 21-cm transition is $2.9 \times 10^{-15}\:\mathrm{s}^{-1}$, equivalent to triplet state lifetime of $10^7$ years \citep{wild52}. Even so, the vast amounts of hydrogen in the Universe lead to 21-cm observations being achievable \citep{vandehulst45}.
The intensity of 21-cm radiation is determined by the spin temperature, $T_{\mathrm{spin}}$, defined through \citep{field58}:

\begin{equation}
\frac{n_1}{n_0}=3 \exp\left(\frac{-T^*}{T_{\mathrm{spin}}}\right)
\end{equation}

\noindent The spin temperature is a fundamental measure of the number densities of the triplet and singlet states ($n_1$, $n_0$) where $T^*=\frac{h\nu_{10}}{k_b}=0.0681\:\mathrm{K}$.

$T_\mathrm{b}$ is detected differentially as a deviation from $T_{\mathrm{CMB}}$, $\delta T_\mathrm{b}$ \citep{field58,field59,ciardi03}:

\begin{eqnarray}
 \delta T_\mathrm{b} &=& 28\: \mathrm{mK} \times (1+\delta)x_{\mathrm{HI}}\left(1-\frac{T_{\mathrm{CMB}}}{T_{\mathrm{spin}}}\right)\left(\frac{\Omega_b h^2}{0.0223}\right) \times \nonumber \\
   && {}\sqrt{\left(\frac{1+z}{10}\right)\left(\frac{0.24}{\Omega_m}\right)}
\end{eqnarray}

\noindent where $\delta$ is the mass density contrast, $h$ is the Hubble constant in units of $100\: \mathrm{km s}^{-1}\:\mathrm{MPc}^{-1}$, $x_{\mathrm{HI}}$ is the fraction of neutral hydrogen and $\Omega_b$ and $\Omega_m$ are the baryon and matter densities in critical density units.

\subsection{Foregrounds}
\label{fg_sim}
Though there have been foreground observations at frequencies relevant to LOFAR using WSRT \citep{bernardi09,bernardi10} the foreground contamination at the frequencies and resolution of LOFAR remains poorly constrained. As a result, foreground models directly relevant for this paper rely on using constraints from observations at different frequency and resolution ranges. These constraints are used to normalize the necessary extrapolations made from observations to create a model relevant for LOFAR-EoR observations.

In general, the foreground components are modelled as power laws in 3+1 dimensions (i.e. three spatial and frequency) such that $T_\mathrm{b} \propto \nu^{\beta}$ \citep[e.g.][]{shaver99,ali08,jelic08,jelic10}.

The foreground simulations used in this paper are obtained using the foreground models described in \citet{jelic08,jelic10}. The foreground contributions considered in these simulations are:

\begin{enumerate}
\item Galactic diffuse synchrotron emission (GDSE) originating from the interaction of free electrons with the Galactic magnetic field. Incorporates both the spatial and frequency variation of $\beta$ by simulating in 3 spatial and 1 frequency dimension before integrating over the $z$-coordinate to get a series of frequency maps. Each line of sight has a slightly different power law. 
\item Galactic localised synchrotron emission originating from supernovae remnants (SNRs). Together with the GDSE, this emission makes up 70 per cent of the total foreground contamination. Two SNRs were randomly placed as discs per 5$^{\circ}$ observing window, with properties such as power law index chosen randomly from the \citet{green06} catalog\footnote{http://www.mrao.cam.ac.uk/surveys/snrs/}.
\item Galactic diffuse free-free emission due to bremsstrahlung radiation in diffuse ionised Galactic gas. This emission contributes only 1 per cent of total foreground contamination, however it still dominates the 21-cm signal. The same method as used for the GDSE is used to obtain maps, however the value of $\beta$ is fixed to -2.15 across the map.
\item Extragalactic foregrounds consisting of contributions from radio galaxies and radio clusters and contributing 27 per cent of the total foreground contamination. The simulated radio galaxies assume a power law and are clustered using a random walk algorithm. The radio clusters have steep power spectra and are based on a cluster catalogue from the Virgo consortium\footnote{http://www.virgo.dur.ac.uk/} and observed mass-luminosity and X-ray-radio luminosity relations.
\end{enumerate} 

Unlike \citet{jelic08,jelic10}, this paper does not consider the polarisation of the foregrounds. The foregrounds simulated here are up to five orders of magnitude larger than the signal we hope to detect. Since interferometers such as LOFAR measure only fluctuations, foreground fluctuations dominate by `only' three orders of magnitude. We will investigate polarized removal in a further analysis.

\subsection{Noise}
\label{noise}
The sensitivity of a receiving system is ultimately determined by the system noise \citep*{thompson01}. The system noise consists of contributions from both the sky and the receivers themselves. Frequency dependence is introduced into the noise through the sky noise and through the frequency dependence of the effective area of the telescope. It is assumed that the noise across an image is Gaussian at any one frequency. For a detailed look at the expected noise on LOFAR measurements, see \citet{labropoulos09}.

We have decided to reproduce the method for calculating the noise here as the noise can have a significant effect on foreground cleaning methods.

\begin{table}
\caption{LOFAR and simulation parameters} 
\begin{tabular}{l c c}
\hline\hline 
Parameter & Description & Value \\ [0.5ex] 
\hline 
$n_d$ & number of dipoles per tile & 16\\ 
$n_t$ & number of tiles per station & 24  \\
$n_s$ & number of stations & 48  \\
$\eta_a$ & antenna efficiency & 1\\
$\eta_s$ & system efficiency & 0.9\\
$\Delta \nu$ & frequency interval [MHz] & 0.5 \\
$t_{\mathrm{int}}$ & integration time [h] & $600$  \\ 
$\Omega_{\mathrm{arcmin}}$ & area of synthesized beam [arcmin$^2$] & 4  \\
$\Omega_{\mathrm{sr}}$ & area of synthesized beam [sr] & $1.35 \times 10^{-6}$ \\ [1ex] 
\hline 
\end{tabular}
\label{table} 
\end{table}

\begin{figure}
\begin{center}
\includegraphics[width=84mm]{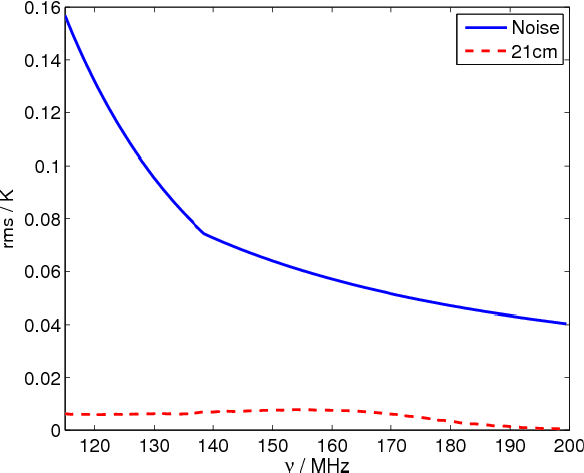}
\caption{The rms of the simulated zero mean noise (blue;solid) and 21-cm (red;dash) maps over frequency.}
\label{nocs_rms}
\end{center}
\end{figure}

 Our parameters for calculating the noise are listed in Table \ref{table}. In order to create a noise curve, Fig. \ref{nocs_rms} we use the following prescription:
  
\noindent The system noise temperature consists of sky brightness and instrumental components. We calculate this system temperature using:
\begin{equation}
T_{\mathrm{sys}} = 140 + 60 \left(\frac{\nu}{300\:\mathrm{MHz}}\right)^{-2.55}
\end{equation}

\noindent The effective area of the array is determined by multiplying the effective area of a single dipole by the number of dipoles in the array where the effective area of a dipole is limited by the size of the tile. We calculate the effective area of the LOFAR array using:
\begin{equation}
A_{\mathrm{eff}} = \mathrm{min}\left(\frac{\lambda^2}{3},1.5625\right) n_d n_t
\end{equation}

\noindent The System Equivalent Flux Density (SEFD) then depends on both of the quantities calculated above:
\begin{equation}
\mathrm{SEFD} = \frac{2 T_{\mathrm{sys}} k_b}{\eta_a A_{\mathrm{eff}}}
\end{equation}

\noindent Finally we calculate the LOFAR noise sensitivity:
\begin{equation}
\sigma = \frac{1}{\eta_s} \frac{\mathrm{SEFD}}{\sqrt{n_s (n_s-1) \Delta \nu t_{\mathrm{int}}}\Omega_{\mathrm{sr}}}
\label{rmsjy}
\end{equation}

Fig. \ref{nocs_rms} shows the noise curve calculated from this prescription compared to the rms of our 21-cm simulation.

For each frequency a LOFAR measurement set was filled with Gaussian noise in the $uv$ plane. This was then imaged to create a real-space image, the root mean square of which can be normalized to the value as given by the prescription described above. For example the noise sensitivity at 150 MHz for an integration time of 600 h and a frequency spacing of 0.5 MHz is 64 mK. The 170 noise maps were uncorrelated over frequency - i.e. a different noise realization was used to fill the measurement set for each frequency.

\subsection{Dirty Images}
\label{beam}
\begin{figure}
\includegraphics[width=84mm]{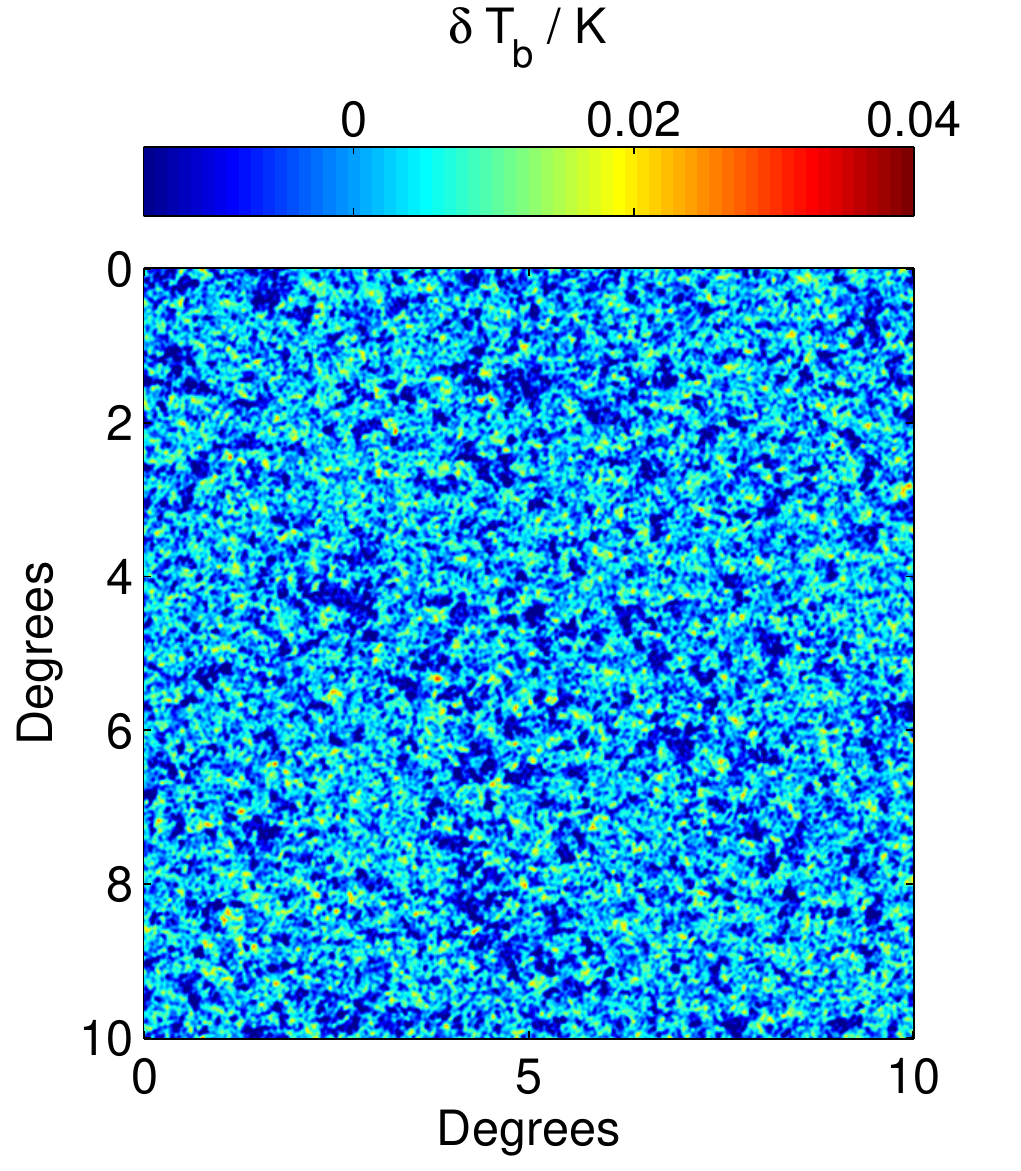}
\caption{The 21-cm signal at 150 MHz, convolved with the PSF. The signal is entirely in emission - this map has been adjusted to have a mean of zero to reflect the observations of an interferometer.}
\label{cs_conv}
\end{figure}

\begin{figure}
\includegraphics[width=84mm]{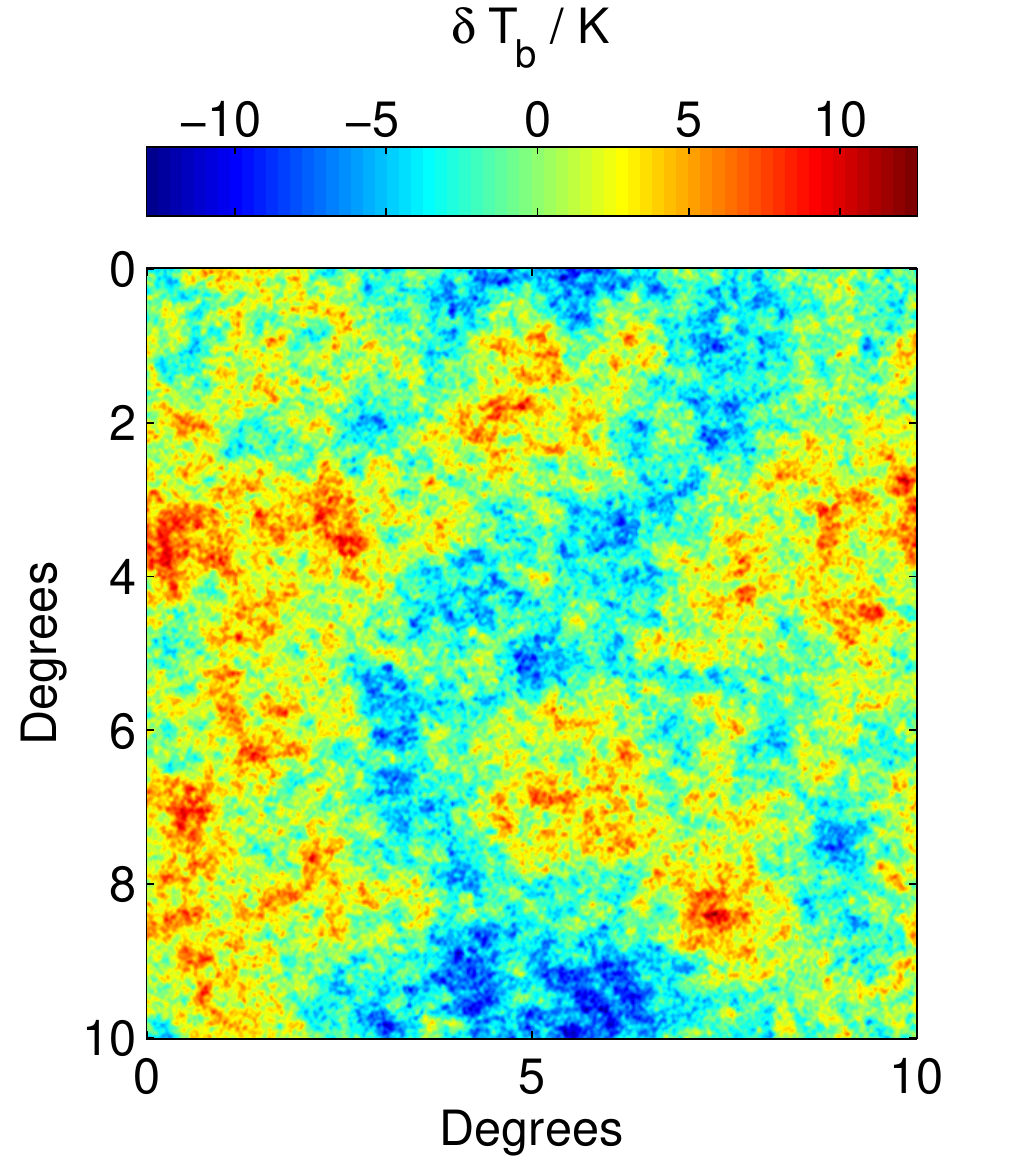}
\caption{The total contribution of the simulated foregrounds at 150 MHz, convolved with the PSF.}
\label{fg_conv}
\end{figure}

The success of an interferometer such as LOFAR is highly dependent on how $uv$ space is sampled. The particular pattern of $uv$ sampling forms a beam which affects how the components such as the foregrounds are seen by the interferometer. Dirty images were simulated by convolving with the PSF of the LOFAR set up used to simulate the noise in the previous section, Fig. \ref{cs_conv} and Fig. \ref{fg_conv}.

The PSF used for creating dirty images (and for creating the noise as described in the previous section) was chosen to be the worst in the observation bandwidth - i.e. the PSF at 115 MHz. In observations the synthesized beam decreases in size with increasing frequency, causing point source signals to oscillate with the beam, producing a foreground signal with an oscillatory signal very much like that of the 21-cm signal. However, this mode-mixing contribution has been found not to threaten the 21-cm recovery and have a power well below the 21-cm level (\citealt{bowman06}; \citealt*{liu09a}). As such we leave the consideration of a frequency dependent PSF to a future paper.

\begin{figure}
\includegraphics[width=84mm]{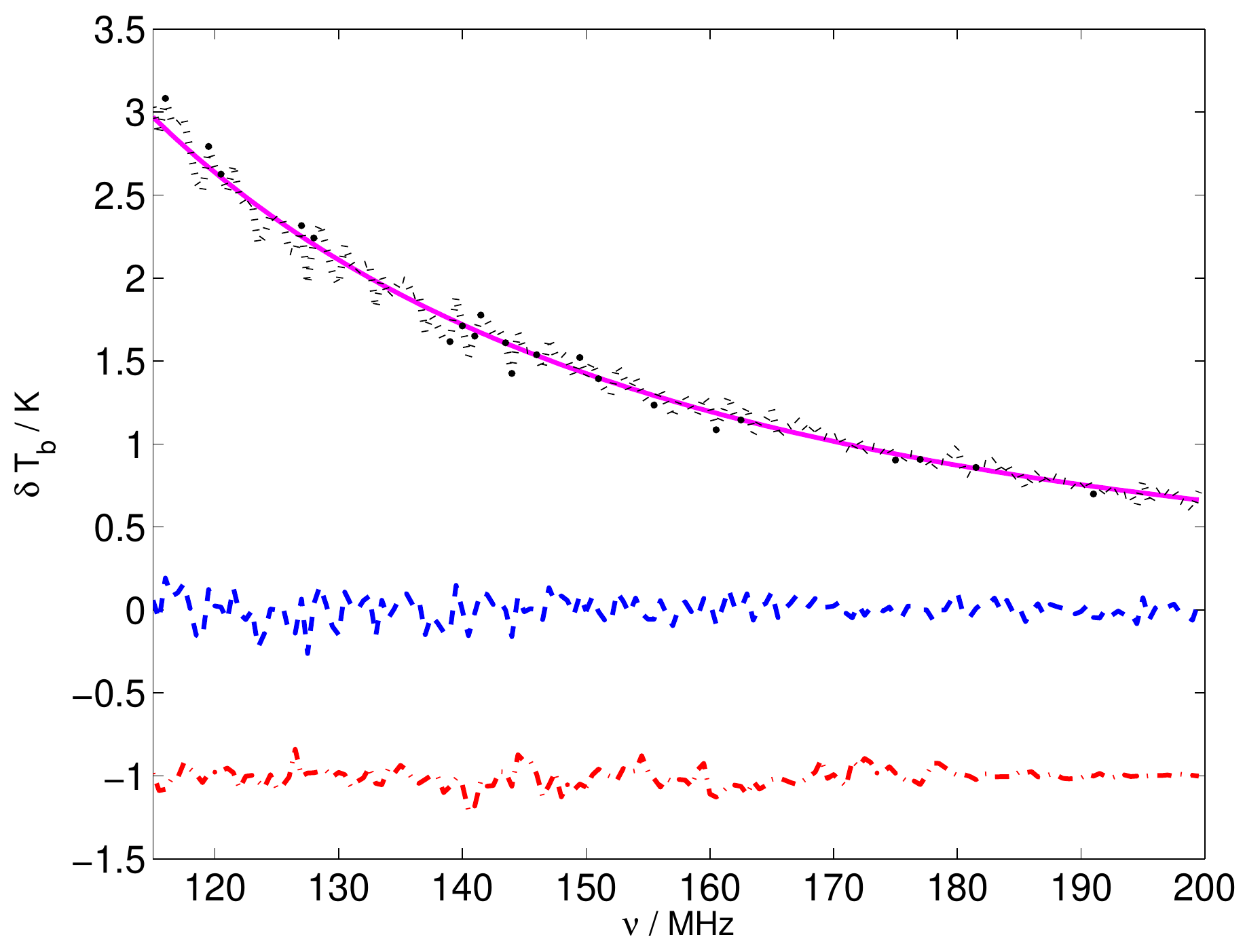}
\caption{The redshift evolution of the simulated cosmological signal (red; dash dot), foregrounds (pink;solid), noise (blue; dash) and total combined signal (black; dot). All components have undergone the PSF convolution. Note the 21-cm signal has been amplified by 10 and displaced by -1K for clarity.}
\label{sims_los}
\end{figure}

Once the foregrounds and 21-cm signal have been adjusted for $uv$ sampling, the three component cubes are added together. The components of the total $\delta T_\mathrm{b}$ along a random line of sight are shown in Fig. \ref{sims_los}.

\subsection{Fourier Transformed Data}
\label{fft}
The \textsc{fastica} method was implemented separately with data both in real and Fourier space. For the latter method, the fiducial image cube was 2D Fourier transformed at each frequency to create a Fourier data cube. The complete cube was then processed with \textsc{fastica} and the output reverse Fourier transformed to obtain the ICs in real space. Unless otherwise stated all results refer to real space implementation.

\section{Results}
\label{results}

\subsection{The Independent Components}
\begin{figure}
\includegraphics[width=84mm]{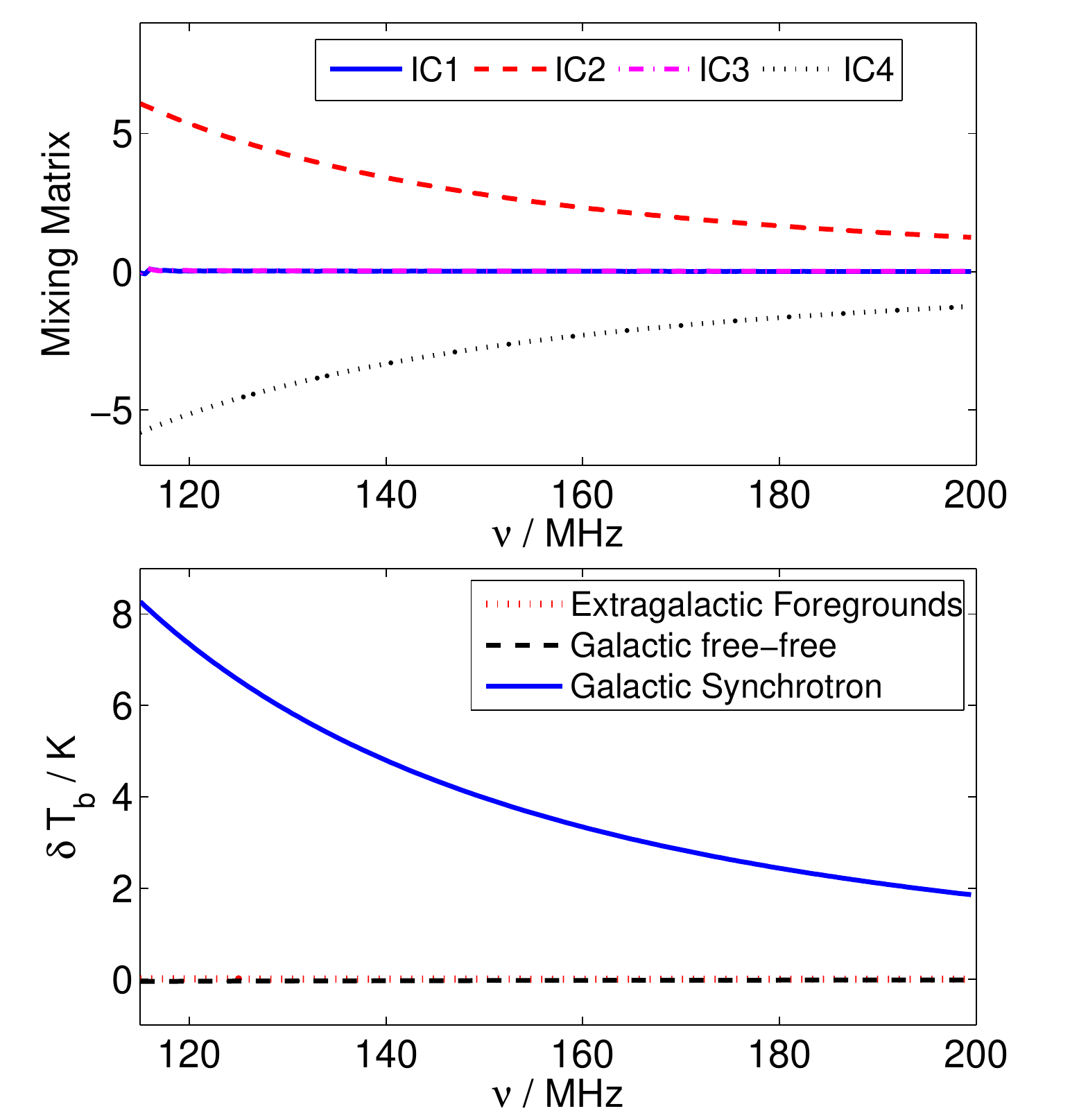}
\caption{In the top panel we show the four columns of the mixing matrix representing the four ICs. The brightness temperatures of the foreground contributions along a random line of sight are shown in the bottom panel. We see that the ICs are each a scaled mixture of the foreground contributions.}
\label{fg_mix}
\end{figure}

The top panel of Fig. \ref{fg_mix} shows the four ICs found by \textsc{fastica} for a clean data cube. These ICs are the columns of the mixing matrix, $\mathbfss{A}$. For comparison we show the line of sight $\delta T_\mathbf{b}$ of the simulated foreground contributions in the bottom panel of Fig. \ref{fg_mix}.

We can see that no single component corresponds to any one single foreground contribution, even when processing a clean data cube. Instead, the components are all a mixture. While in ICs 2 and 4 the presence of Galactic synchrotron is obvious, in the other components the combination of the contributions is not so clear. It is also worth noting that while IC4 shows a significant contribution from Galactic synchrotron, it is inverted. \textsc{fastica} can only determine the ICs up to a multiplicative constant and so the sign and magnitude of the components are irrelevant.

\begin{figure}
\includegraphics[width=84mm]{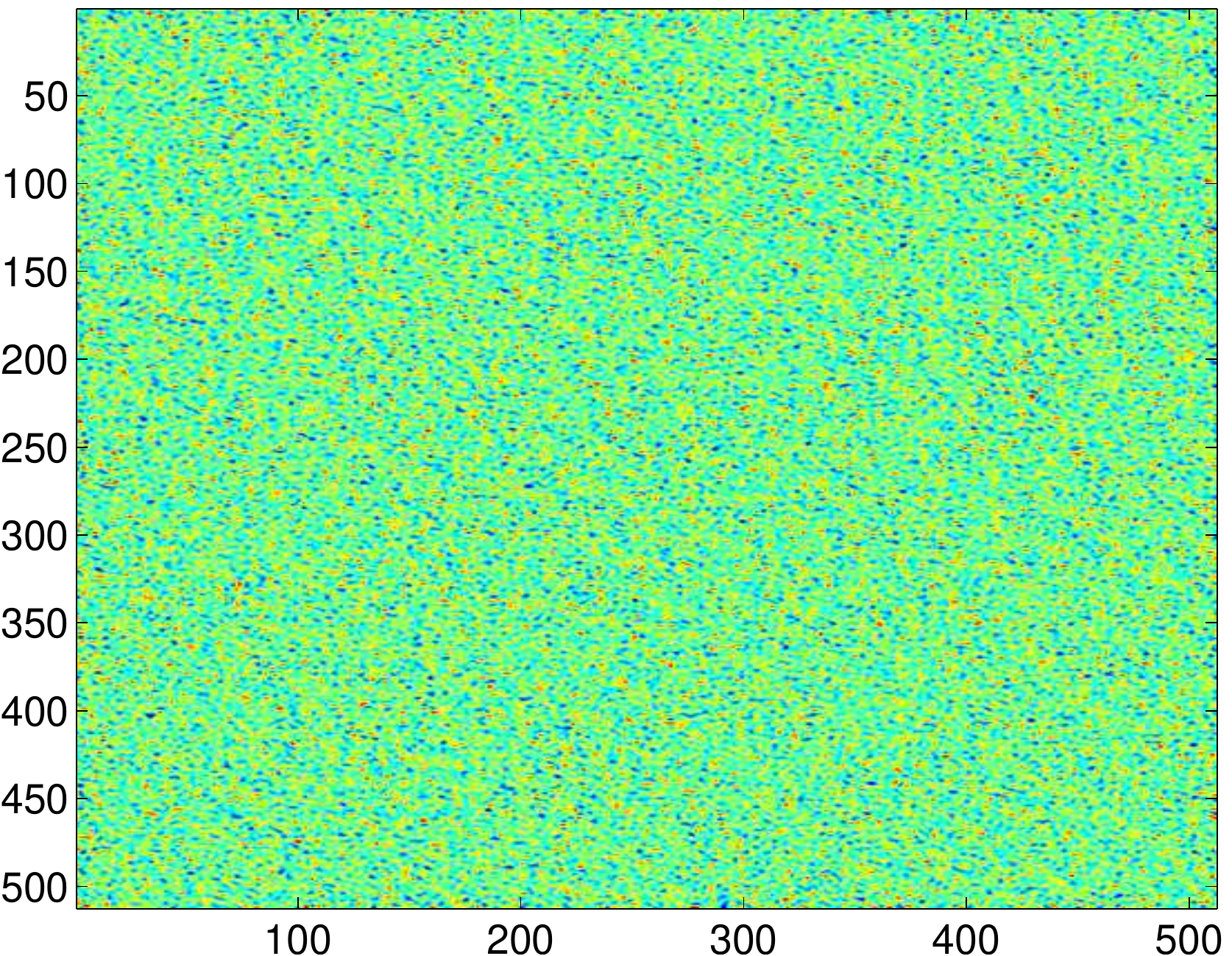}
\caption{The first coefficient map of the ICs when \textsc{fastica} processes the clean data cube.}
\label{IC1}
\end{figure}

\begin{figure}
\includegraphics[width=84mm]{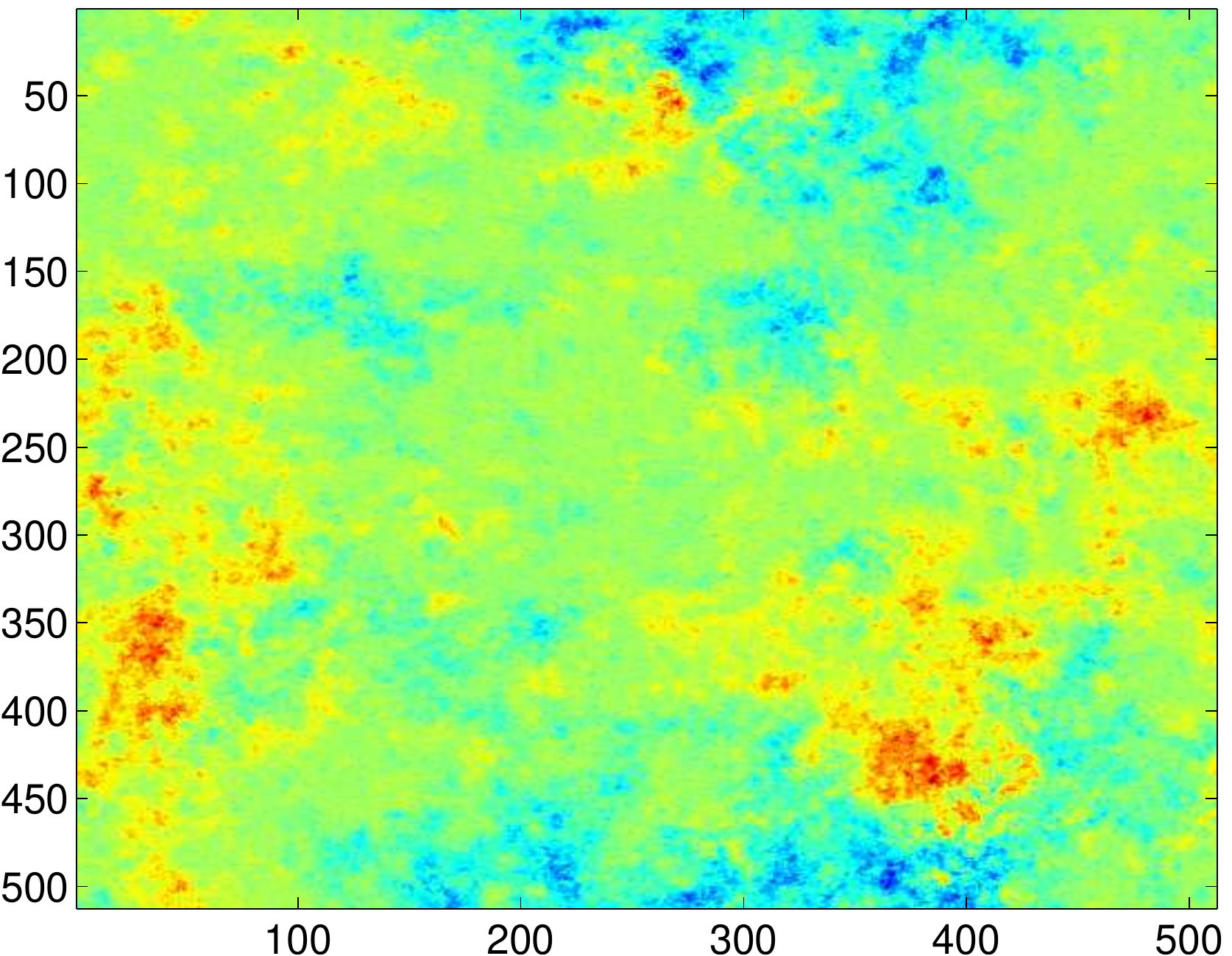}
\caption{The second coefficient map of the ICs when \textsc{fastica} processes the clean data cube.}
\label{IC2}
\end{figure}

\begin{figure}
\includegraphics[width=84mm]{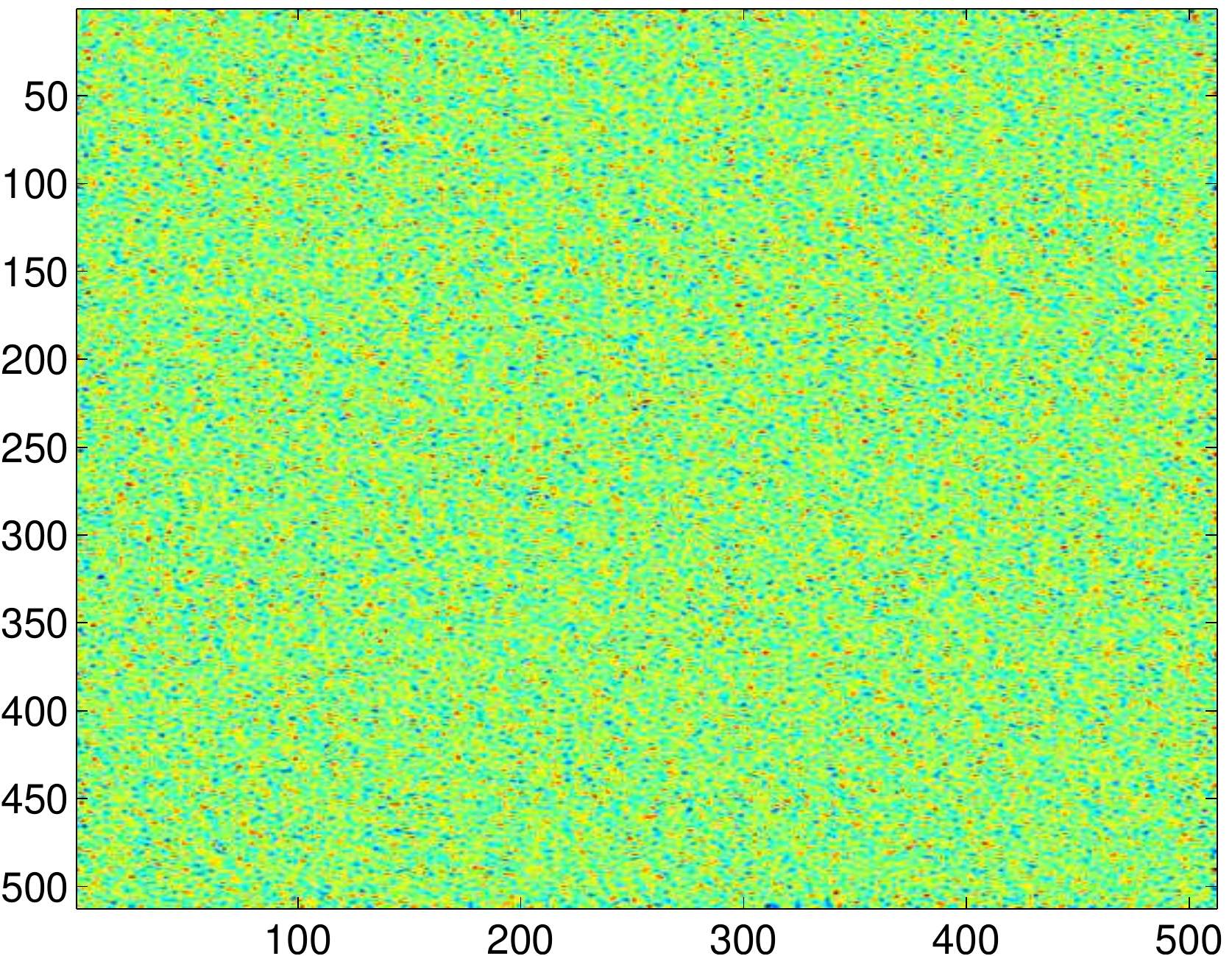}
\caption{The third coefficient map of the ICs when \textsc{fastica} processes the clean data cube.}
\label{IC3}
\end{figure}

\begin{figure}
\includegraphics[width=84mm]{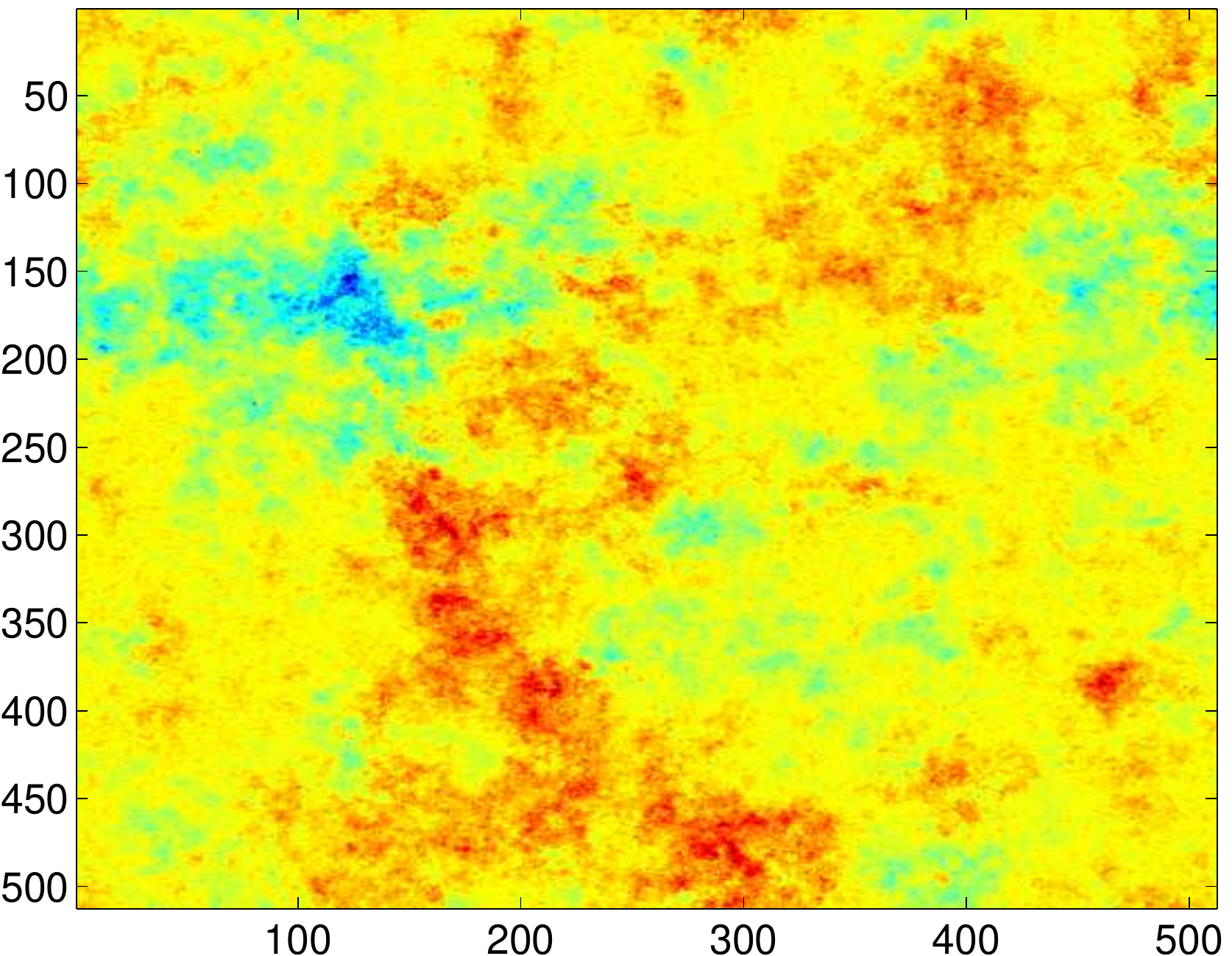}
\caption{The fourth coefficient map of the ICs when \textsc{fastica} processes the clean data cube.}
\label{IC4}
\end{figure}

\begin{figure}
\includegraphics[width=84mm]{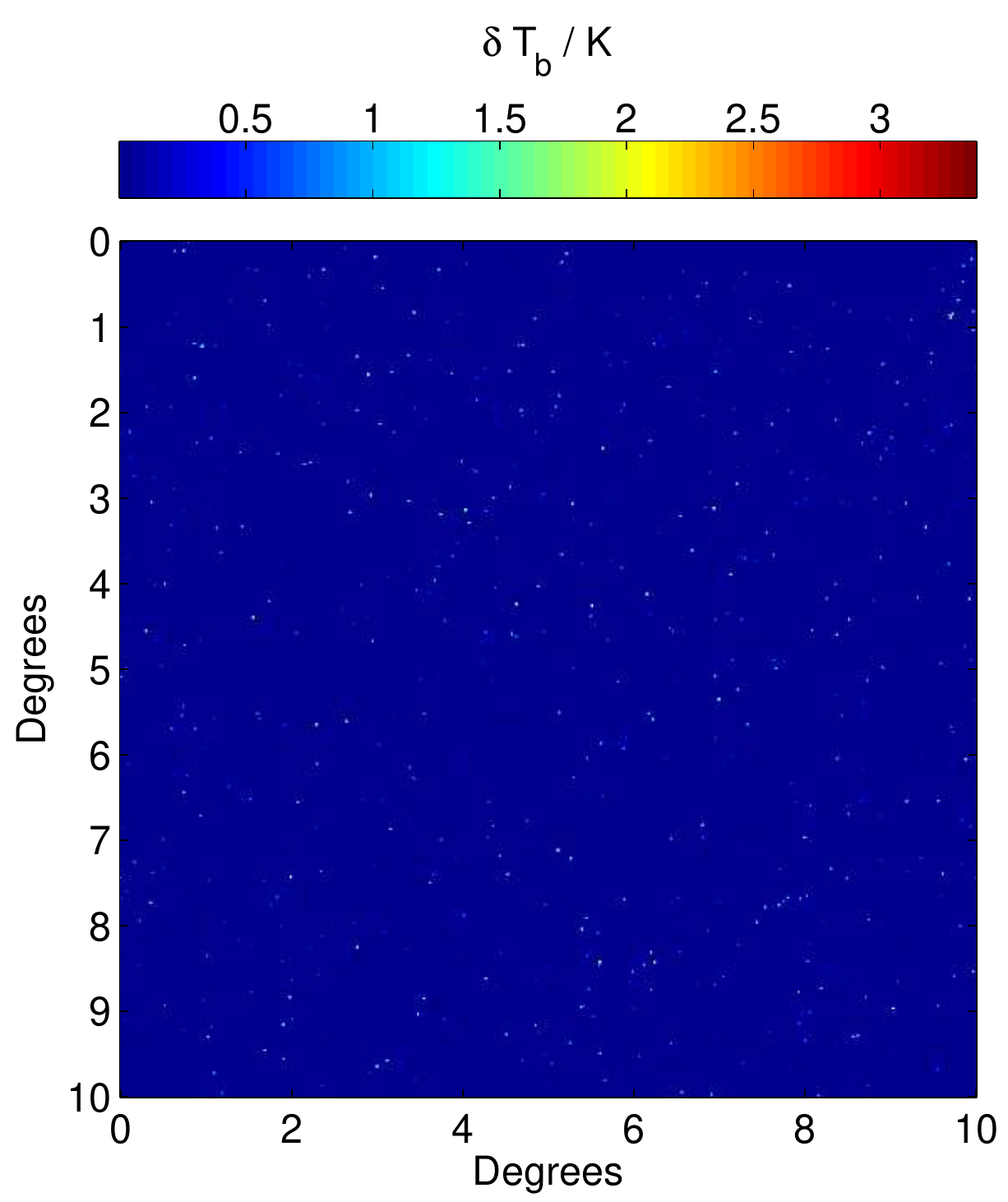}
\caption{The simulated extragalactic foregrounds at 150 MHz.}
\label{EGfg}
\end{figure}

\begin{figure}
\includegraphics[width=84mm]{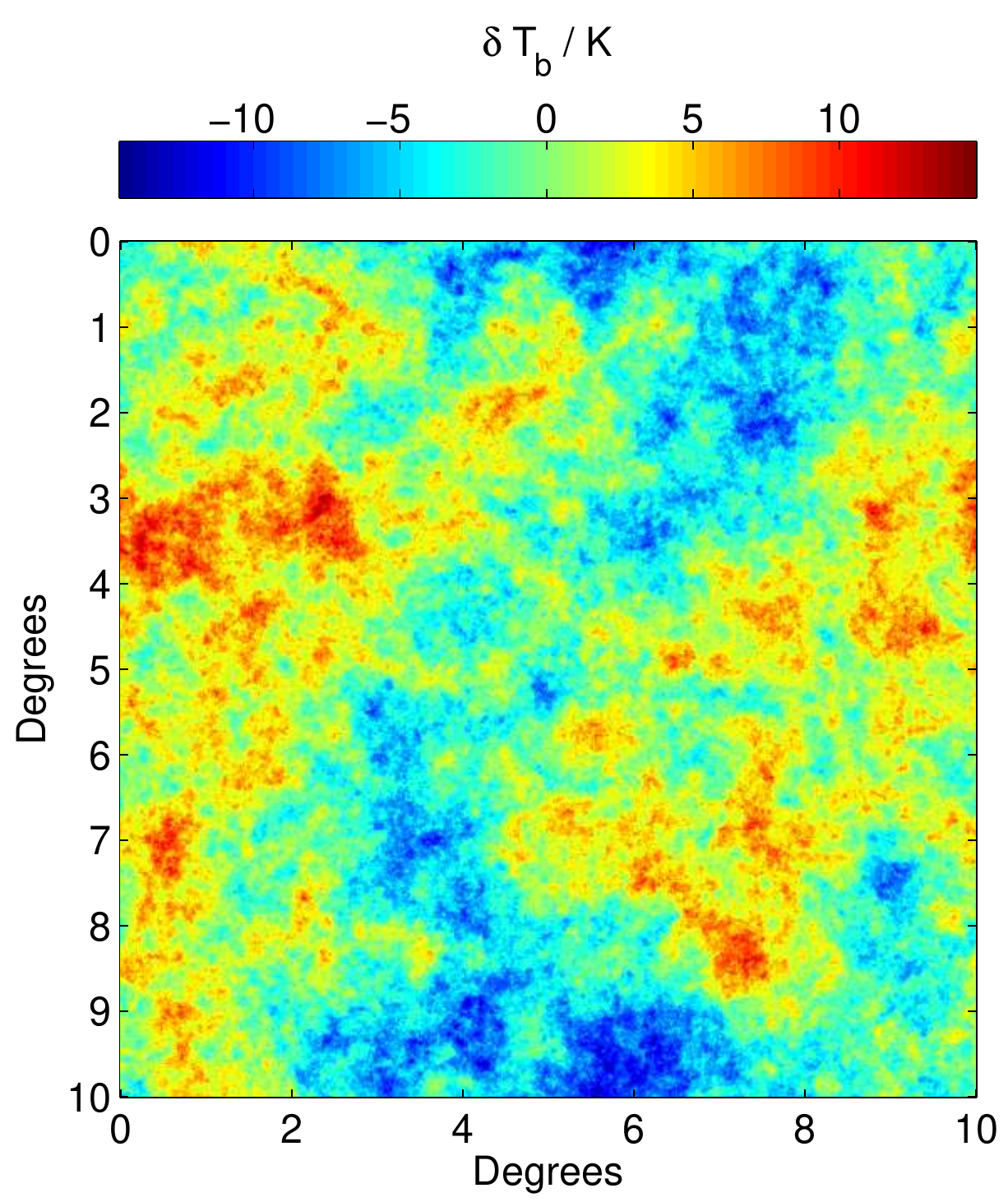}
\caption{The simulated Galactic synchrotron foregrounds at 150 MHz.}
\label{Gsyn}
\end{figure}

\begin{figure}
\includegraphics[width=84mm]{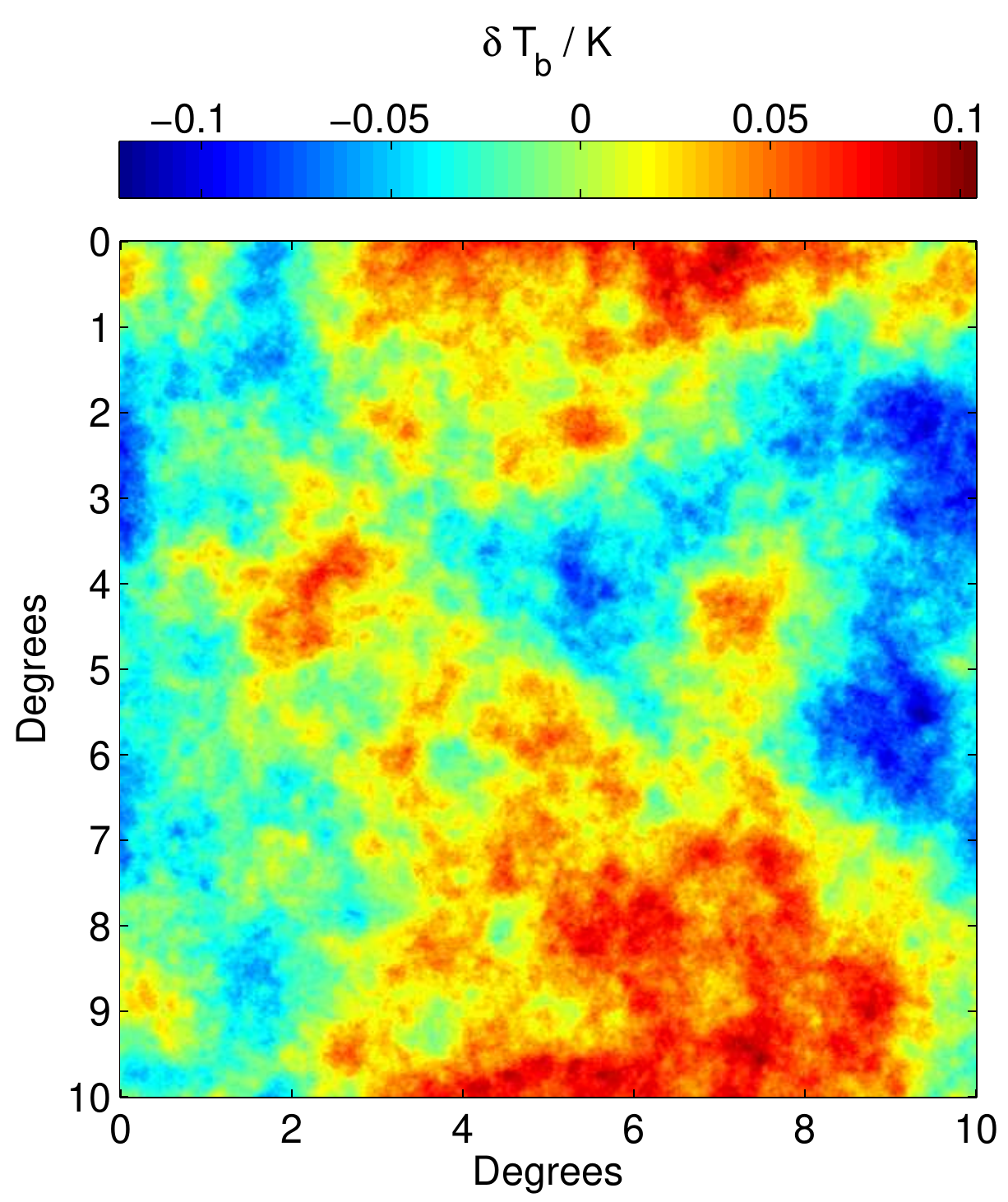}
\caption{The simulated Galactic free-free foregrounds at 150 MHz.}
\label{Gff}
\end{figure}

The coefficients of the ICs are stored in the matrix $\mathbfss{s}$ and are presented in Fig. \ref{IC1}, Fig. \ref{IC2}, Fig. \ref{IC3} and Fig. \ref{IC4}. We can compare these coefficients to the maps of the foreground contributions, Fig. \ref{EGfg}, Fig. \ref{Gsyn} and Fig. \ref{Gff}. We see that all four coefficients are a mixture of the contributions as expected.

\subsection{Fitting Errors and Variance}
We will first discuss the \textsc{fastica} results on the simulation where the data cube has been convolved with the PSF, the data processing is carried out in real space and four ICs are assumed. The word `simulated' is used to refer to the input maps and `reconstructed' is used for the estimates resulting from \textsc{fastica}. The total input signal is separated into reconstructed foregrounds and residuals. The residuals are the difference between the original mixed signal and the reconstructed foregrounds.

To evaluate the accuracy of the foreground fitting by \textsc{fastica}, we calculated the foreground fitting error, Equation \ref{fgfit}, for each pixel. 

\begin{equation}
\mbox{fitting error} = \frac{\mathrm{fg_{reconstructed}}-\mathrm{fg_{simulated}}}{\mathrm{fg_{simulated}}} \times 100.0
\label{fgfit}
\end{equation}

\begin{figure}
\includegraphics[width=84mm]{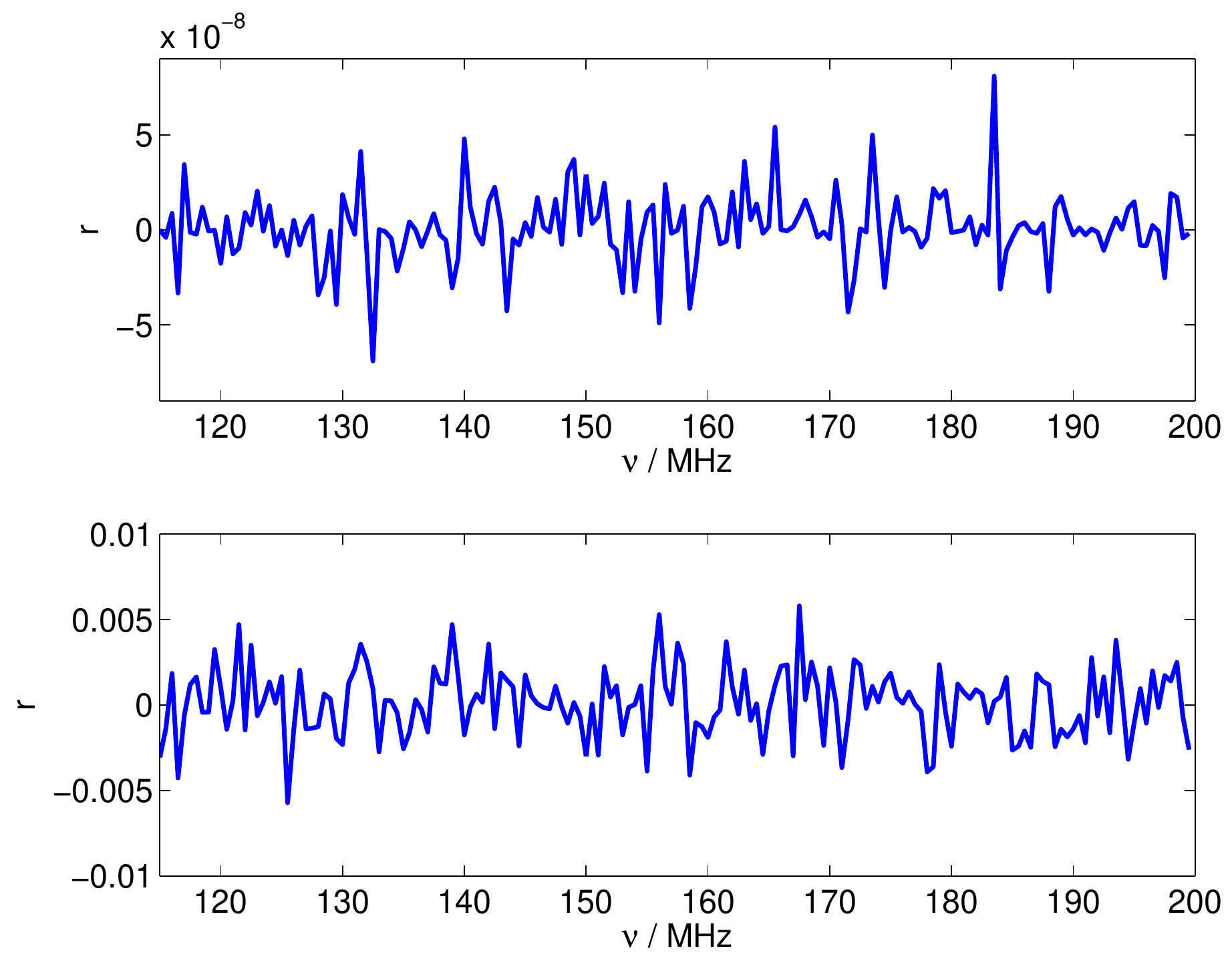}
\caption{a) The Pearson correlation coefficient between the foreground maps and foreground fitting errors. b) The Pearson correlation coefficient between the noise maps and foreground fitting errors.}
\label{pear_nofg}
\end{figure}

In Fig. \ref{pear_nofg} we plot the Pearson correlation coefficient between the foreground fitting errors and foregrounds (top) and between the foreground fitting errors and the noise (bottom). The Pearson correlation coefficient between two data sets $a$ and $b$ is defined as:
\begin{equation}
r = \frac{\sum_i (a_i-\bar{a})(b_i - \bar{b})}{\left[\sum_i (a_i-\bar{a})^2 \sum_i (b_i - \bar{b})^2\right]^{\frac{1}{2}}}
\end{equation}

\noindent where $\bar{a}$ is the mean of the data set $a_i$, $\bar{b}$ is the mean of the data set $b_i$ and the measure is normalized such that $r=\pm 1$ for correlation/anti-correlation.

We see that there is very little correlation between the foreground maps and the foreground fitting errors, with around six magnitudes more correlation between the noise maps and the foreground fitting errors.

To get a representation of the foreground fitting error over an entire map, the rms error of the fitted foregrounds was calculated, Fig. \ref{fg_rms}. It should be noted that this error takes into account all scales - including those with a disproportionate error as will be seen in the power spectra. The rms difference between the simulated and reconstructed foregrounds was calculated over all $512^2$ lines of sight for each frequency. Also, an rms error for each map was calculated using only 68 $\%$ of the pixels - with the pixels of lowest error selected first.  When the outlier pixels are discounted we find that the rms error is below 10 mK for the majority of the frequency range. This is still high enough to be of concern as the 21-cm signal is itself of order tens of mK, however the inclusion of all scales means this is a worst case scenario. 

\begin{figure}
\includegraphics[width=84mm]{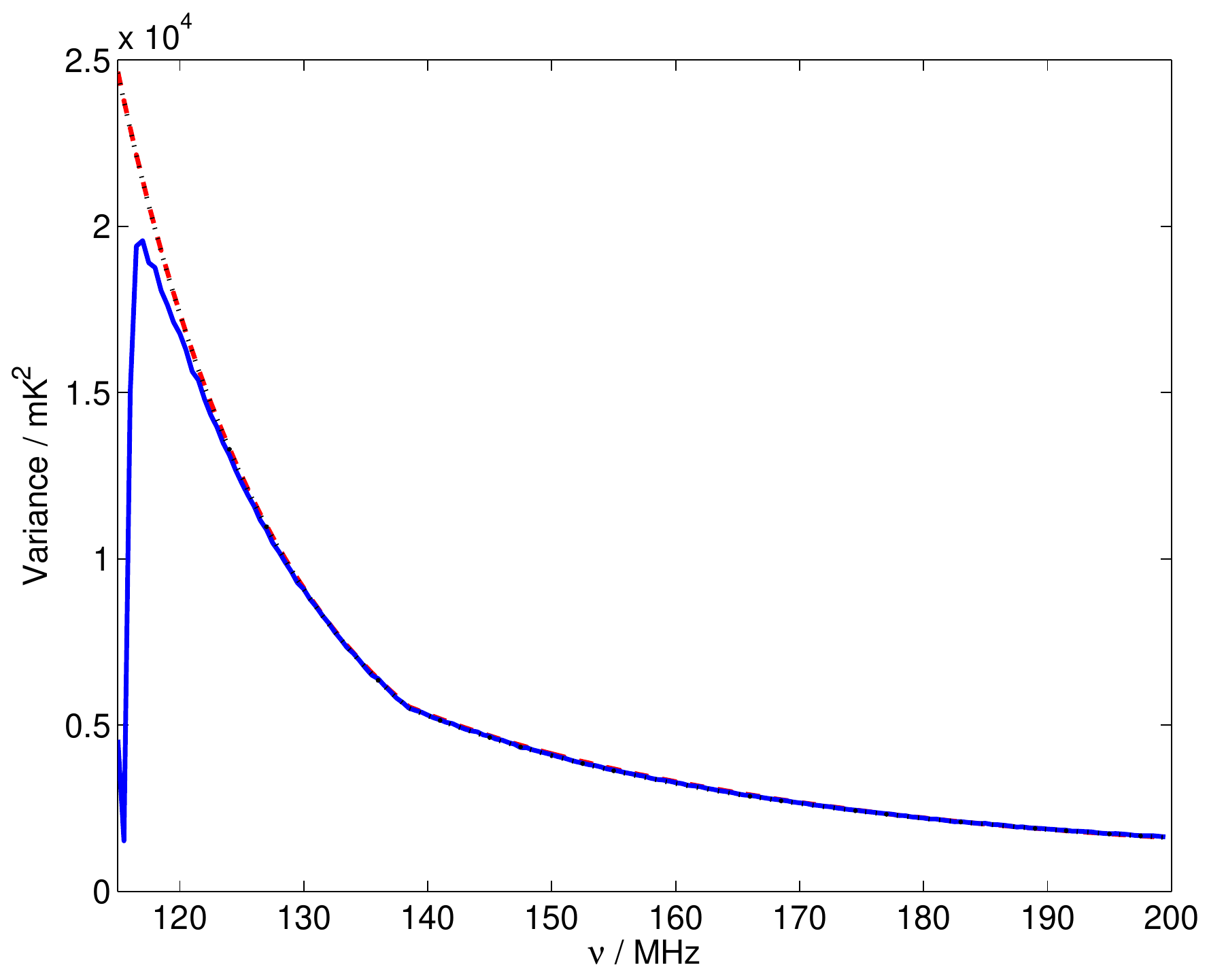}
\caption{The variance across the combined simulated cosmological signal and noise (red, dash), noise alone (black, dot) and residuals (blue, solid).}
\label{rest_fid}
\end{figure}

For a statistical detection of the Epoch of Reionization, LOFAR aims to detect a non-zero variance after the noise and foregrounds have been accounted for. We begin by combining the simulated noise and simulated 21-cm signal and taking the variance of this signal. This can then be compared to the variance of the \textsc{fastica} residuals - Fig. \ref{rest_fid}. The residual variance is recovered at all but the smallest frequencies. At frequencies below 120 MHz (or $z>10.8$) the variance is significantly underestimated, probably as a result of foreground overfitting - the leakage of noise power into the estimated foreground power. This failure at very low frequencies is hardly surprising considering that this is where the noise and foregrounds are at their strongest.
 
We subtract the variance of the simulated noise directly from the variance of these residuals: var(reconstructed 21-cm) = var(residuals) - var(noise). This is a fair assumption as we should be able to look at the data in narrow frequency bins and estimate the statistics (e.g. variance and power spectrum) of the noise to a very high accuracy. 

\begin{figure}
\includegraphics[width=84mm]{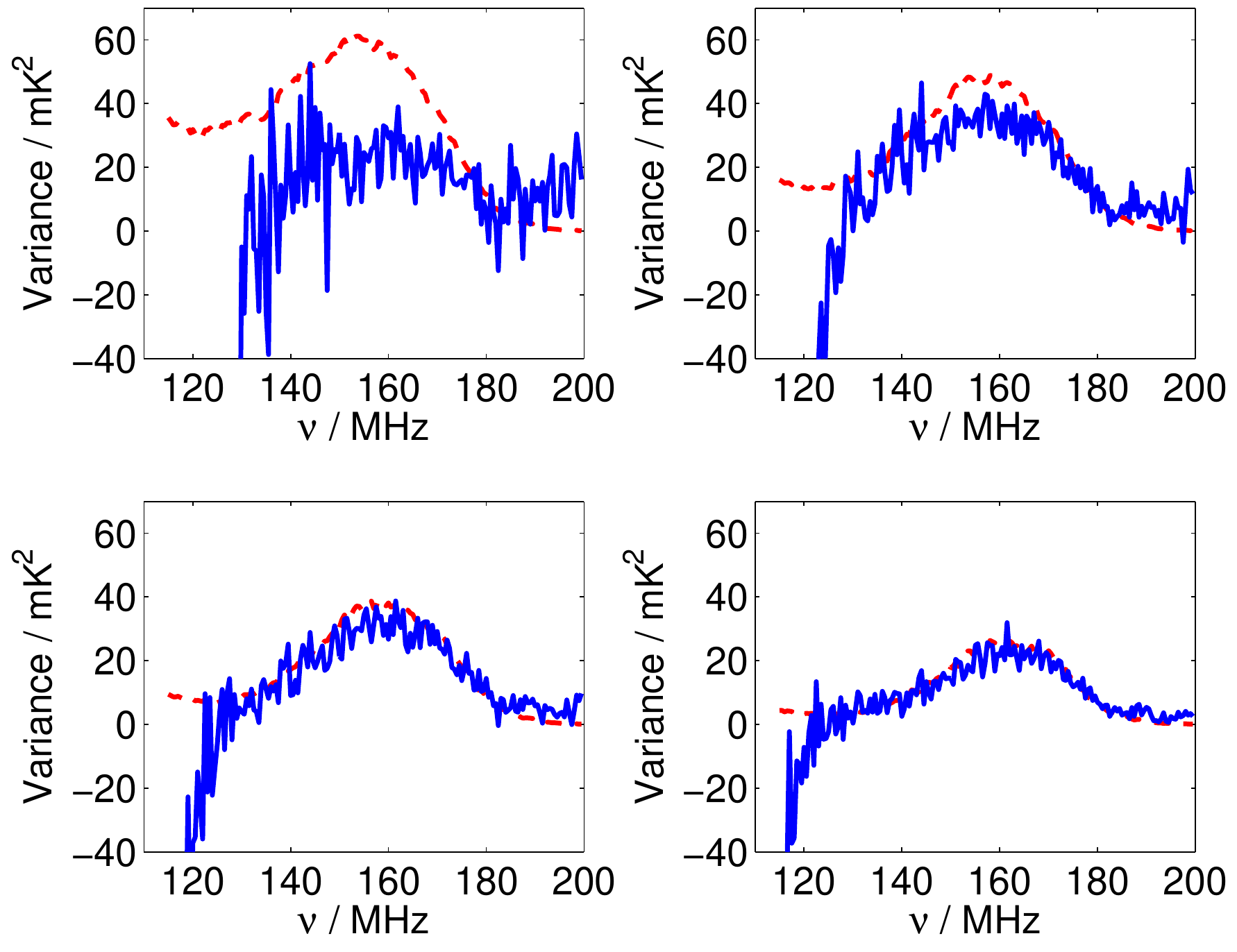}
\caption{The variance across the simulated (red; dash) and reconstructed 21-cm maps (blue; solid) for the fiducial data and data which has had Fourier filtering of modes below 2,3 and 5 PSF scales (in reading order).}
\label{var_fid}
\end{figure}

\begin{figure}
\includegraphics[width=84mm]{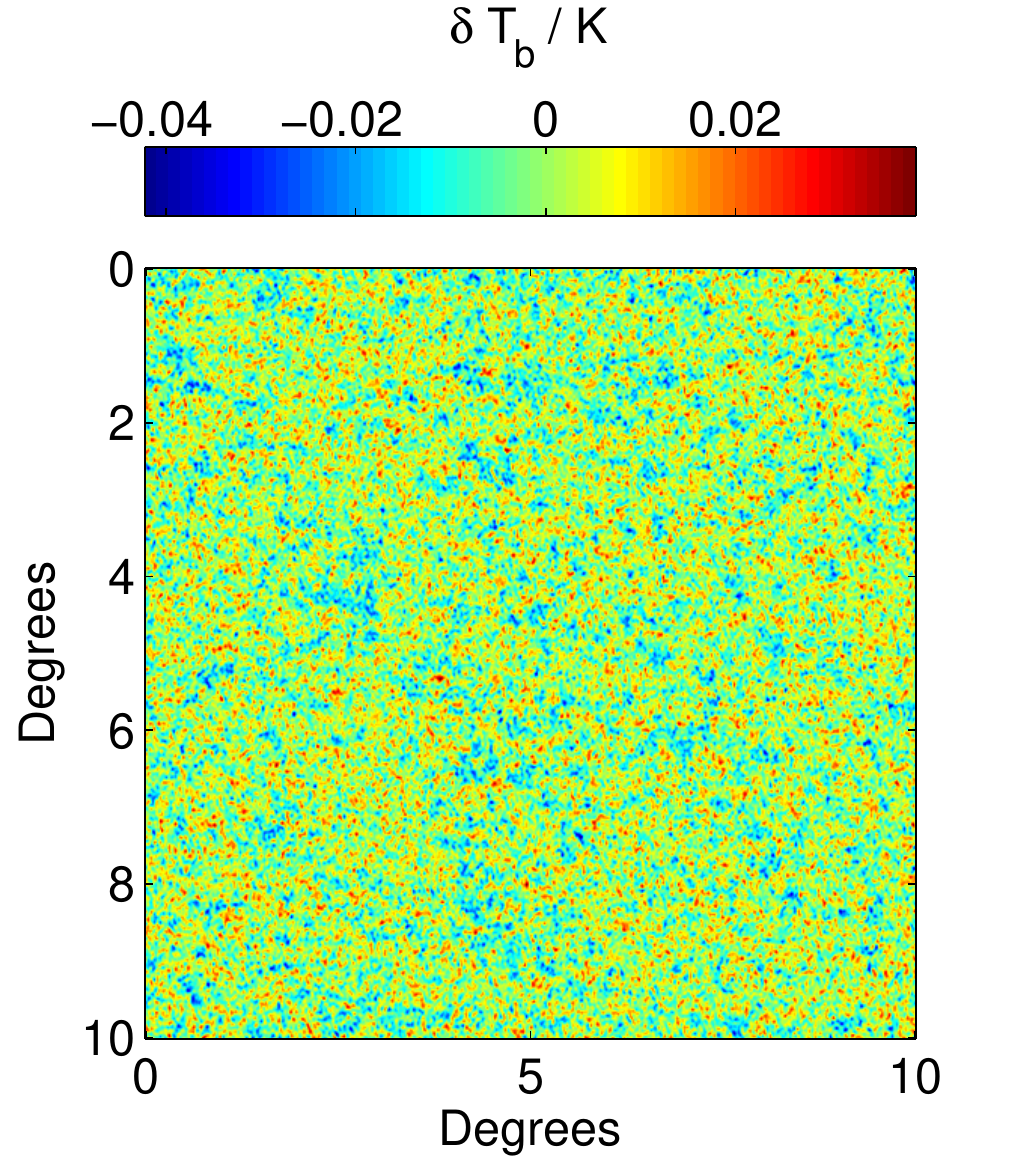}
\caption{The reconstructed 21-cm signal at 150 MHz for dirty data. We see that while there is a strong correlation between the large scale structure in this image and the original signal, Fig. \ref{cs_conv}, there is also a large amount of excess small scale structure, probably due to noise leakage into the foregrounds.}
\label{csrec_150}
\end{figure}

We find that the recovered 21-cm variance, Fig. \ref{var_fid} top-left, is not robust to small scale power in the original signal. By removing the noise simulation maps manually from the residual maps in order to get crude maps of the recovered 21-cm signal, excess small scale power is evident, Fig. \ref{csrec_150}. We note that we do this direct noise subtraction for a crude visual inspection only and not for any of the analytical results. The excess power is most likely due to \textsc{fastica} not being robust to the small scale power (noise) in our data, allowing it to leak into the reconstructed foregrounds. It was found that by Fourier filtering the data to entirely remove $k$ modes below a threshold corresponding to a multiple of the PSF scale, the variance recovery was significantly improved. A very good recovery occurs with filtering below 5 PSF scales (Fig. \ref{var_fid} bottom-right).

At the extremes of the frequency range the reconstructed variance increasingly diverges from that of the simulated 21-cm. Both the noise and the foregrounds are at their largest at lower frequencies meaning that both fitting errors and noise leakage is likely to be largest here, leading to less accurate 21-cm reconstruction. Equally, at the larger frequencies, the 21-cm signal is almost non-existent making an accurate reconstruction difficult when swamped with fitting errors and noise. 

This variance calculation was also carried out on a data cube where the residual and noise maps were smoothed from a $512^2$ grid to a $256^2$ grid before the same variance calculation was carried out above and compared to the variance of a smoothed simulated 21-cm map. The curves are, as expected, slightly smoother however the trend and conclusions remain the same.

\subsubsection{Varying the Number of ICs}
\begin{figure}
\includegraphics[width=84mm]{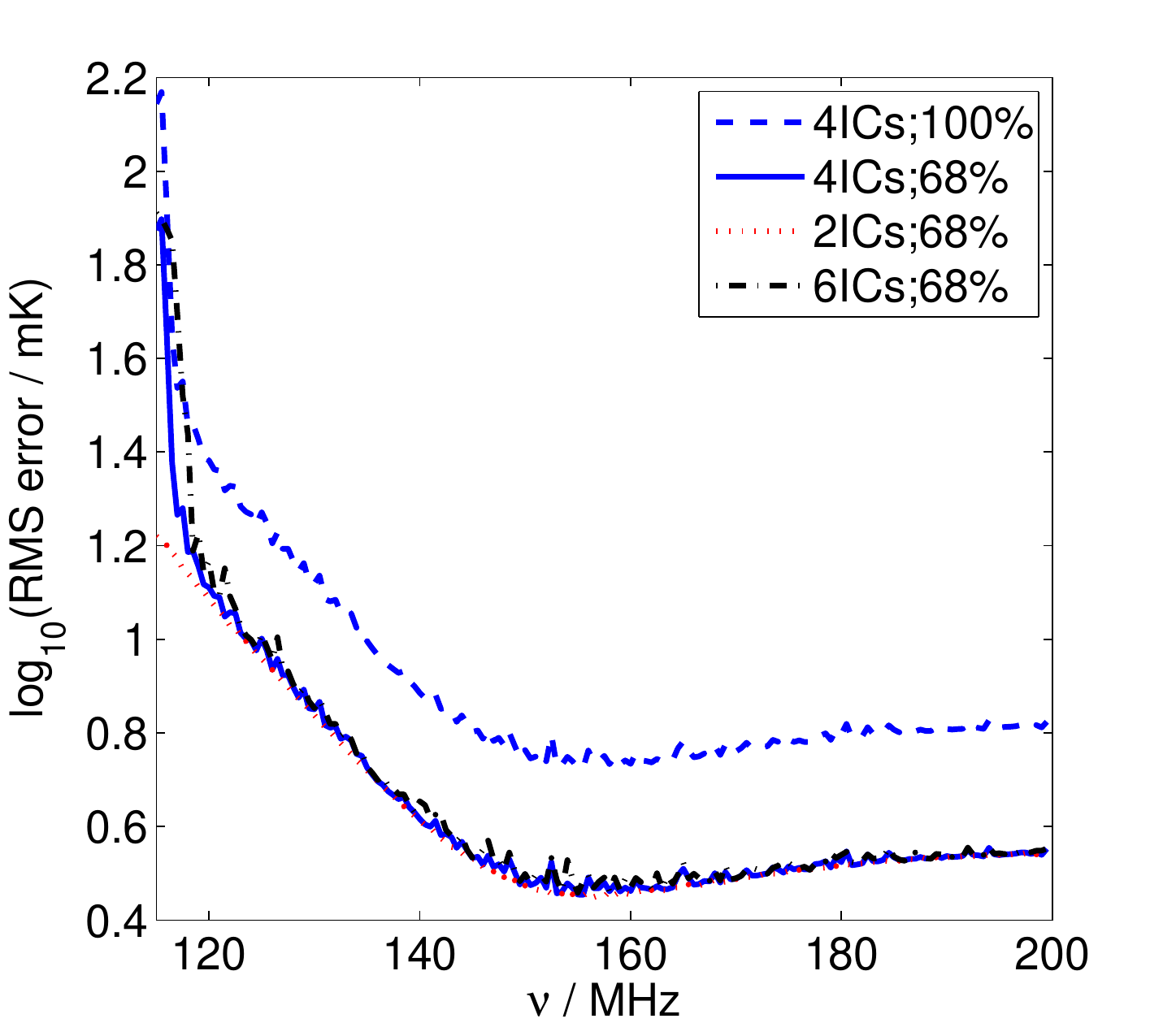}
\caption{The rms error of the 4 IC reconstructed foregrounds for when all pixels are considered (blue;dash) and when only the middle 68 per cent of the error distribution is included (blue;solid). Also, the rms errors of the reconstructed foregrounds for \textsc{fastica} applied according to models with 2 (red; dot) and 6 (black; dashdot) ICs, with only the middle 68 per cent of the error distribution included.}
\label{fg_rms}
\end{figure}

\begin{figure}
\includegraphics[width=84mm]{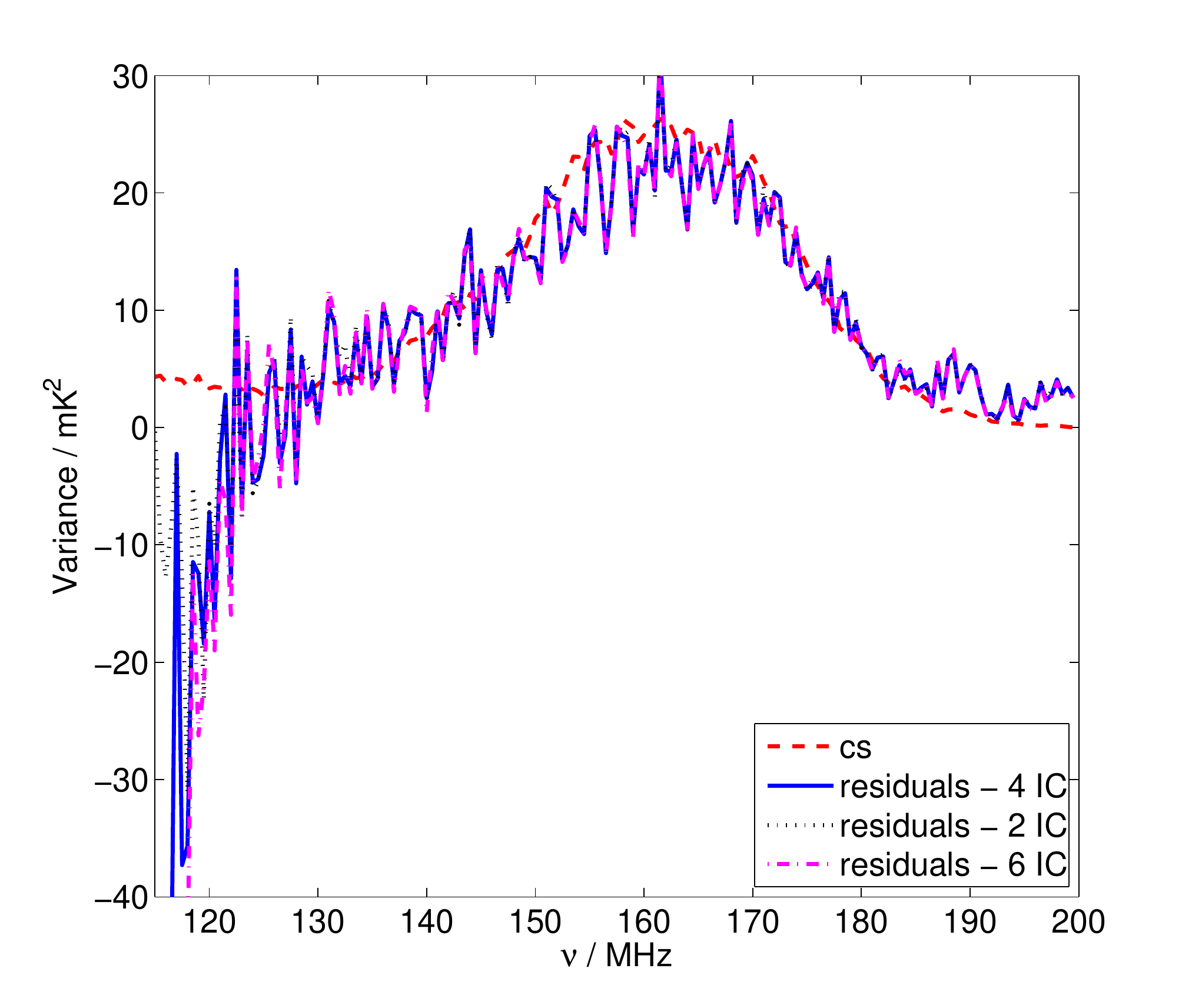}
\caption{The variance across the simulated (red; dash) and reconstructed maps at each frequency, for the \textsc{fastica} algorithm run with the assumption of 2 (black; dot), 4 (blue; solid) and 6 (pink; dot dash) ICs. The data has been Fourier filtered at the 5 PSF scale.}
\label{var_ICs}
\end{figure}

The \textsc{fastica} algorithm requires specification of the number of ICs to be used in the reconstructed foreground model. Though we have modelled the various foreground contributions, it is not a trivial task to determine how these depend on each other and to what degree. To test the sensitivity of our results to the number of ICs chosen we calculate the rms error and variance recovery for IC numbers of 2, 4 and 6 in Fig. \ref{fg_rms} and Fig. \ref{var_ICs}. 

We see that small variations in the number of ICs does not endanger the statistical recovery of the 21-cm signal. For the remainder of this paper, four ICs are assumed.

\subsection{Power Spectra}

Together with the variance, EoR experiments aim to recover the power spectrum of the cosmological signal over a broad range of frequencies.

Different effects are important for modes parallel and perpendicular to the line of sight. For example, consider the scenario where the foregrounds have been under-fitted by a constant over the frequency range. This offset will not be evident in the 1D power spectrum of the residuals, however will be evident in the angular power spectrum if that constant is dependent on line of sight. Thus it has been argued by \citet{harker10} that for LOFAR data, the separate calculation of 1D and 2D power spectra has its advantages. However this does not consider modes neither parallel or perpendicular to the line of sight and as such we calculate 2D and 3D power spectra. We note here that we only performed one simulation of the cosmological signal so the power spectrum error bars relate to this specific realisation of the density field.

\subsubsection{Angular Power Spectra}
The angular power spectrum of a map at a single frequency is calculated by 2D Fourier transforming that map and binning the pixels according to Fourier scale, $k$. The power at any particular $k$, $\langle\delta(k) \delta^*(k)\rangle$ is the average power of all the $uv$ cells in the bin centering on $k$. The error on the point for a particular bin, $k_i$ are calculated as $\alpha_i = \frac{\langle\delta(k_i) \delta^*(k_i)\rangle}{\sqrt{n_{k_i}}}$ where $n_{k_i}$ is the number of $uv$ cells that reside in that $k$ bin. The power spectra were averaged over frequency bandwidths of 2.5 MHz and all frequencies quoted are the middle frequency of the bandwidth. The power spectrum of the reconstructed cosmological signal is calculated by subtraction of the noise power spectrum from the \textsc{fastica} residuals power spectrum. The error on the simulated 21-cm power spectrum is added in quadrature with the error on the noise to reflect the error on the reconstructed 21-cm power spectrum. Note that we assume Gaussianity whereas the 21cm signal is not Gaussian and also we calculate the error bars from the power of a single realization rather than over an ensemble of simulations. We ask the reader to bear in mind that these error bars might be considered incomplete because of this. 

The quantity actually plotted is $\Delta^2_{2D}(k) = \frac{A k^2 \langle\delta(k) \delta^*(k)\rangle}{2 \mathrm{\pi}}$ where $A$ is the area of the simulation map.

\begin{figure}
\includegraphics[width=84mm]{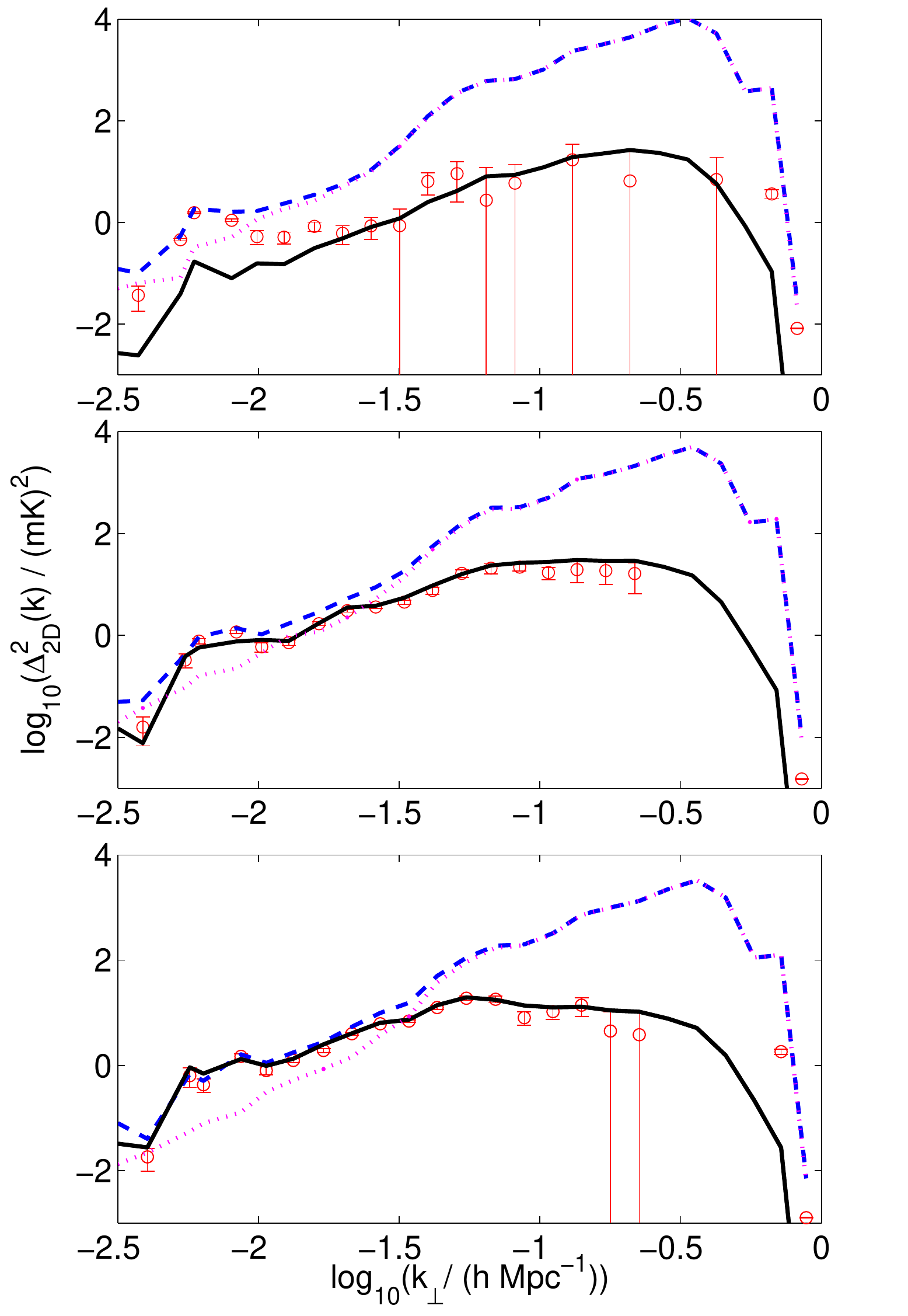}
\caption{2D power spectrum of the simulated 21-cm signal (black;solid), reconstructed 21-cm signal (red;points), residuals (blue,dash) and noise (pink,dotted) at 131 MHz, or $z$=9.84, 151 MHz, or $z$=8.40 and 171 MHz, or $z$=7.30 from top to bottom. Any error bars extending to below the x axis in linear space are shown extending to negative infinity in log space.}
\label{2D_ps}
\end{figure}

Fig. \ref{2D_ps} shows the extent to which the \textsc{fastica} method can recover the 21-cm angular power spectrum. Overall, the 21-cm power spectrum is convincingly recovered across the redshift range. Any points where the power of the residuals are below the power of the noise are omitted, as this leads to an unrealistic negative reconstructed 21-cm power. As such, there is a lack of data at small scales indicative of noise leakage into the foregrounds. 

This noise leakage could be a resolution effect (i.e. artefacts originating from correlated noise) or simply a result of \textsc{fastica} not being robust to noise, most likely the latter. 

\subsubsection{3D Power Spectra}
To calculate the 3D power spectra we divide the cube into sub-bands of 8 MHz to avoid signal evolution effects. For each sub band we then carry out a 3D Fourier transform and calculate the 3D power spectrum in spherical annuli in Fourier space. The frequencies attached to the plots correspond to the centre of the sub band plotted. What we actually plot is the quantity $\Delta^2_{\mathrm{3D}}(k) = \frac{V k^3 \langle\delta(k) \delta^*(k)\rangle}{2 \mathrm{\pi}^2}$ where V is the volume.

\begin{figure}
\includegraphics[width=84mm]{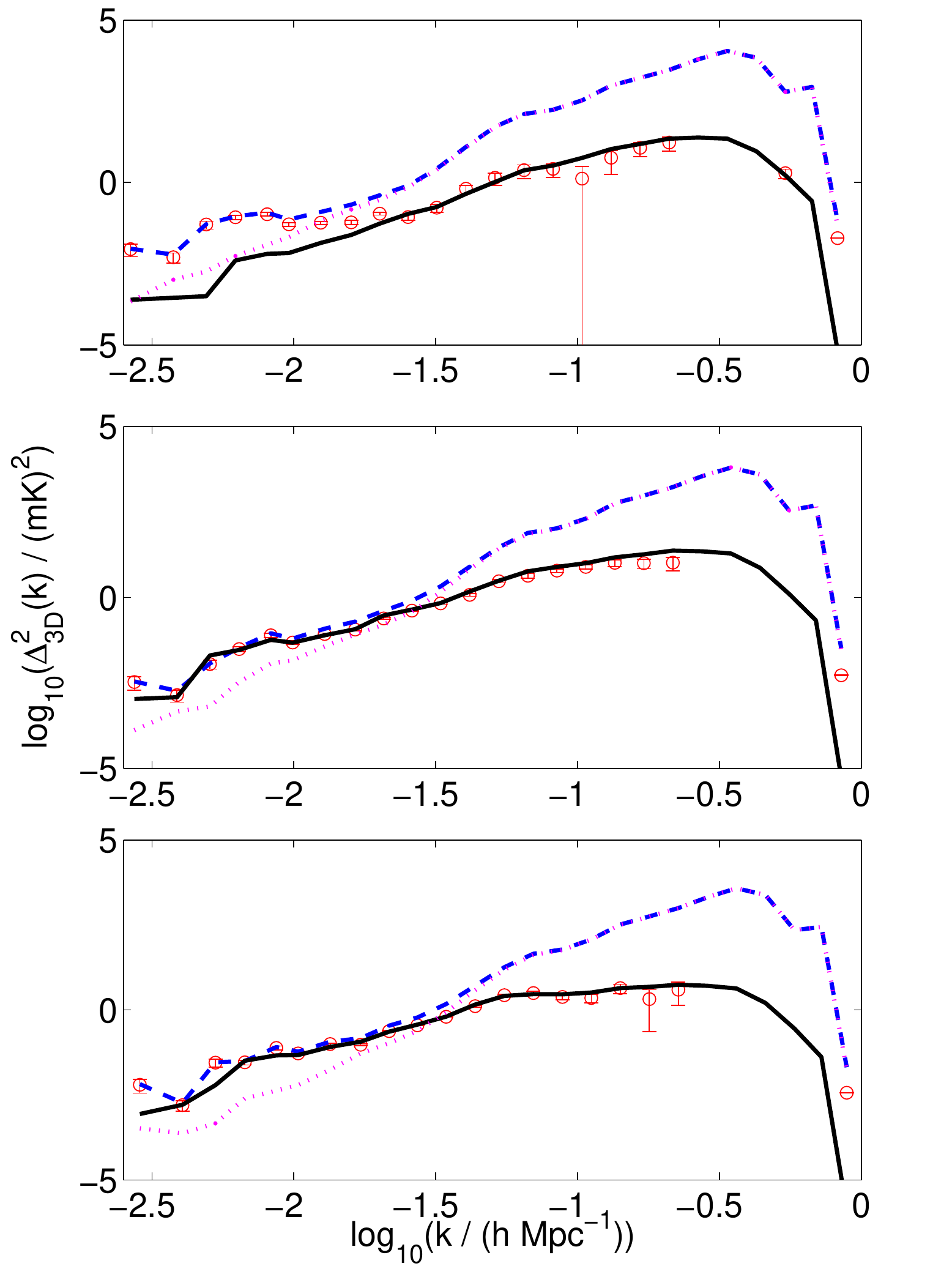}
\caption{3D power spectrum of the simulated 21-cm signal, reconstructed 21-cm signal, residuals and noise at 135 MHz, or z=9.51, 151 MHz, or z=8.40 and 175 MHz, or z=7.11 over an 8 MHz sub band. Any points where the power of the residuals are below the power of the noise are omitted, as this leads to an unrealistic negative reconstructed 21-cm power. The error bars and linestyles are as described in Fig. \ref{2D_ps}.}
\label{3D_ps}
\end{figure}

We find the same accurate recovery on scales above a few multiples of the PSF but with smaller errors due to the larger amount of data evaluated, Fig. \ref{3D_ps}. 

\subsubsection{Cross Correlation Power Spectra}

\begin{figure}
\includegraphics[width=84mm]{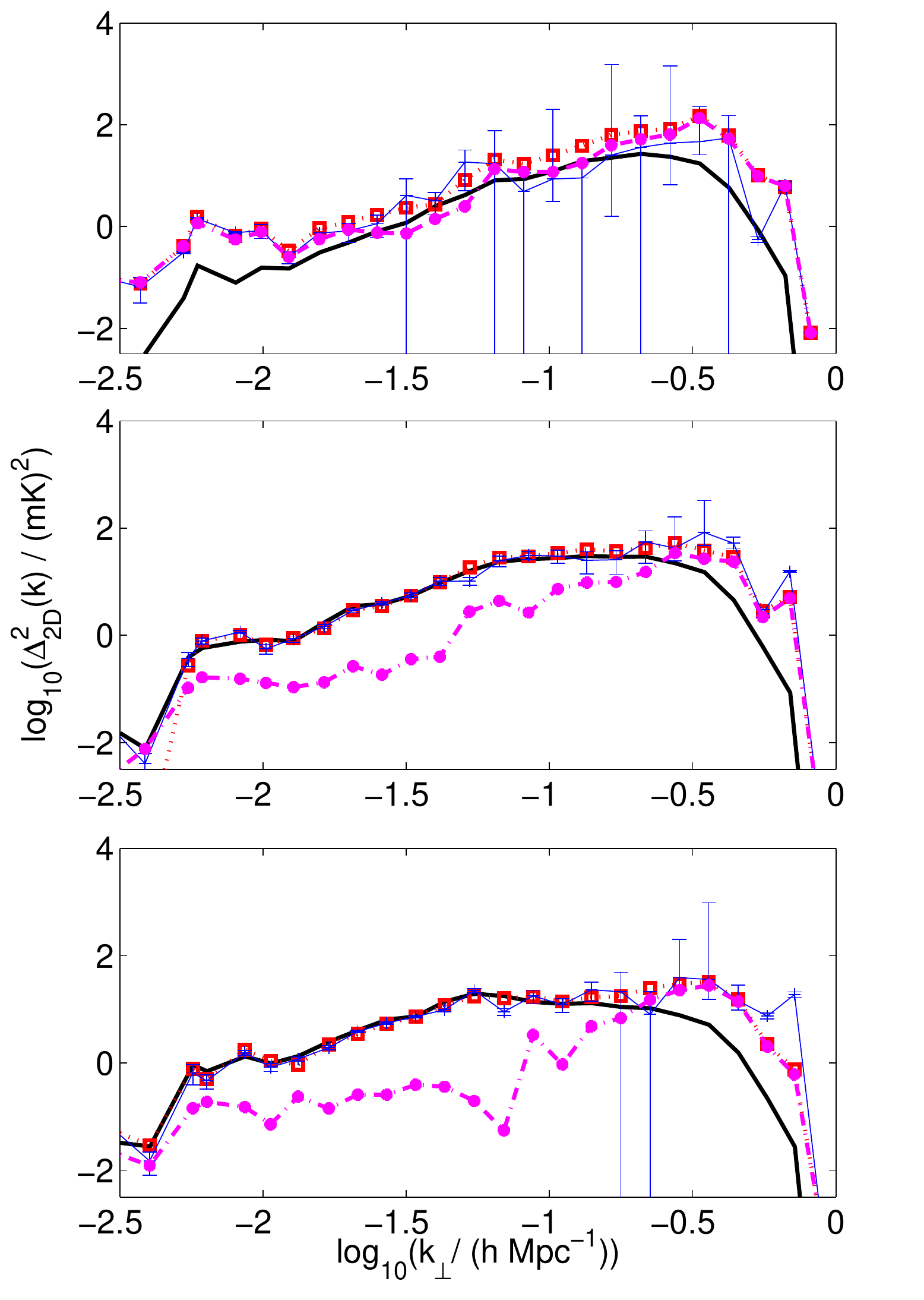}
\caption{Cross correlations of the two residuals (blue,cross), two reconstructed 21-cm signals (red,square), two fitting error estimates (pink,circle) and the auto-correlation of the simulated (black,solid) at 131 MHz, 151MHz and 171MHz. Only one set of error bars is shown for clarity.}
\label{cs_1}
\end{figure}

To try and retrieve a more robust reconstructed 21-cm power spectrum, the cross correlation of two data cube realisations was carried out. Two independent noise realisations were created and combined with identical foregrounds and 21-cm signals to create two data cubes with the only difference being the noise realisation. \textsc{fastica} was performed on both of these cubes separately, resulting in two residual files. We carried out cross correlations on the two reconstructed cosmological signals, the two residual files and the two 21-cm fitting error estimates (i.e. reconstructed 21-cm minus the simulated 21-cm). By cross correlating the two residual signals consisting of two different noise realizations, we increase the amount of noise that will drop out in the noise cross terms, hopefully resulting in a more accurate power spectrum recovery when applied to real data. However we do not expect a significant improvement in comparison to our 2D autospectra here as we have already assumed a perfect knowledge of the noise spectrum. Instead we do this as an example of a more robust method of power spectrum recovery for real data. Note that since correlations can be negative, it is the absolute value that is plotted. The errors bars on the cross spectra are calculated in the same way as for the auto spectra, namely: $\alpha_i = \frac{\langle\delta(k_i) \delta^*(k_i)\rangle}{\sqrt{n_{k_i}}}$ where $n_{k_i}$ is the number of pixels that resided in that k bin. The power spectra recovered as a result of this process are shown in Fig. \ref{cs_1}. 

\begin{figure}
\includegraphics[width=84mm]{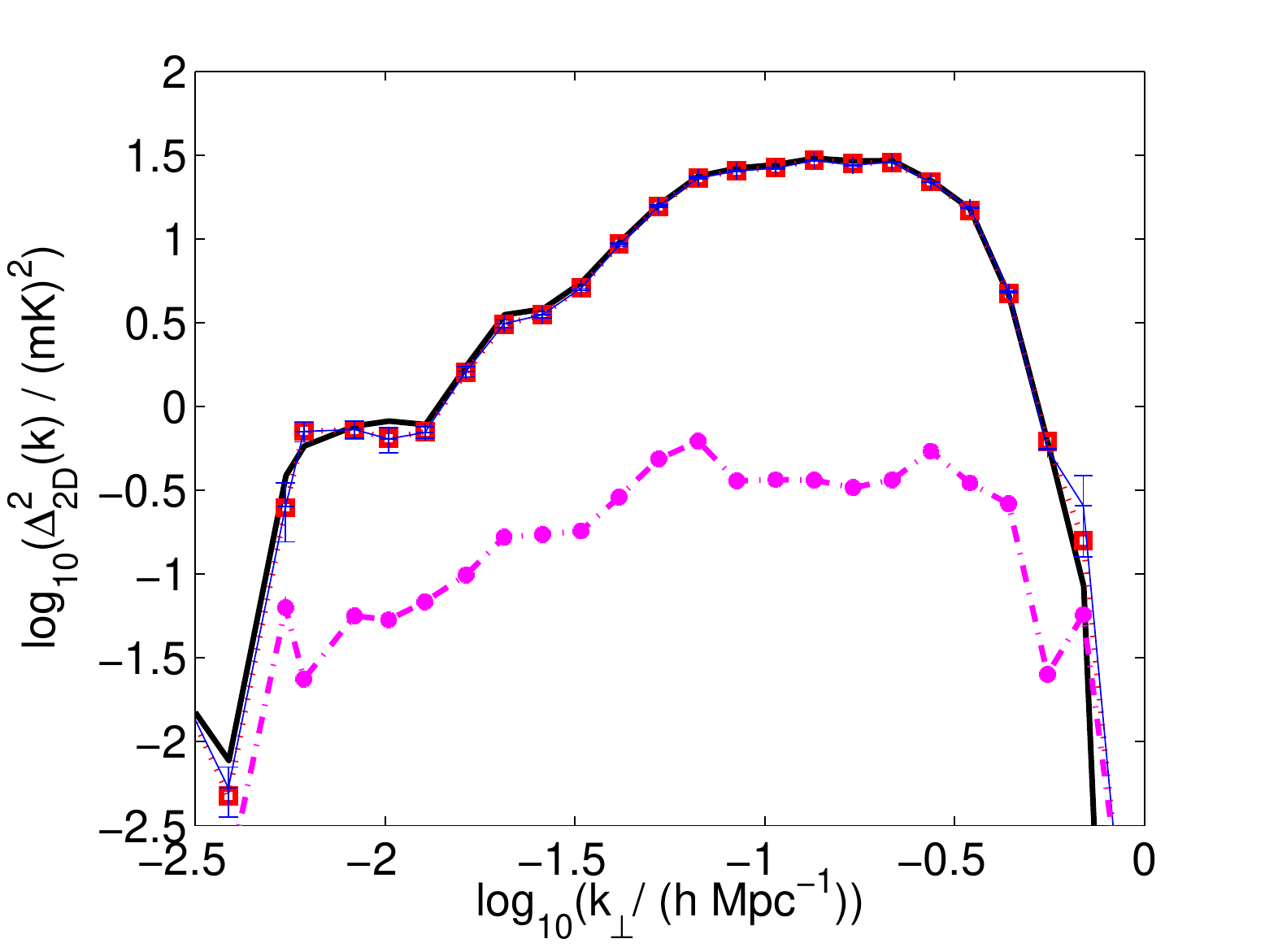}
\caption{Cross correlations of the two residuals (blue,cross), two reconstructed 21-cm signals (red,square), two fitting error estimates (pink,circle) and the auto-correlation of the simulated (black,solid) and reconstructed cosmological signal (for one realization) (red,circles) at 150 MHz. The noise realizations involved have been adjusted to be 10 times smaller than the LOFAR realizations.}
\label{cc_ska}
\end{figure}

The cross correlations were also carried out on two noise realizations which were adjusted to have roughly 10 times the signal to noise ratio of the LOFAR realizations (similar to what is hoped for SKA), Fig. \ref{cc_ska}. We see that with a higher signal to noise ratio the auto and cross correlation estimates are significantly improved.

\subsection{Kurtosis and Skewness}
Skewness and, to a lesser extent, kurtosis have both been suggested as alternative statistics for the 21-cm detection due to their increased robustness to fitting errors compared to the variance \citep{harker09a,wyithe07}. We define skewness ($\gamma_1$) and kurtosis ($\gamma_2$) in Equations \ref{skew} and \ref{kurt} respectively.

\begin{equation}
\gamma_1=\frac{\frac{1}{N} \sum_i (T_i-\bar{T})^3}{(\frac{1}{N} \sum_i(T_i-\bar{T})^2)^{\frac{3}{2}}}
\label{skew}
\end{equation}

\begin{equation}
\gamma_2=\frac{\frac{1}{N} \sum_i (T_i-\bar{T})^4}{(\frac{1}{N} \sum_i(T_i-\bar{T})^2)^{2}}-3
\label{kurt}
\end{equation}

\noindent Kurtosis is defined in such a way that a Gaussian distribution would have a kurtosis of zero.

 The structure of the 21-cm skewness and kurtosis for different source models has been discussed by \citet{harker09a,wyithe07,iliev11}. There is expected to be a skewness of the 21-cm signal as the signal becomes increasingly non-Gaussian as the regions of ionised hydrogen become more numerous. Simulations also show an increase in skewness at very low redshift due to a high brightness temperature tail related to regions with some remaining neutral hydrogen.

\begin{figure}
\includegraphics[width=84mm]{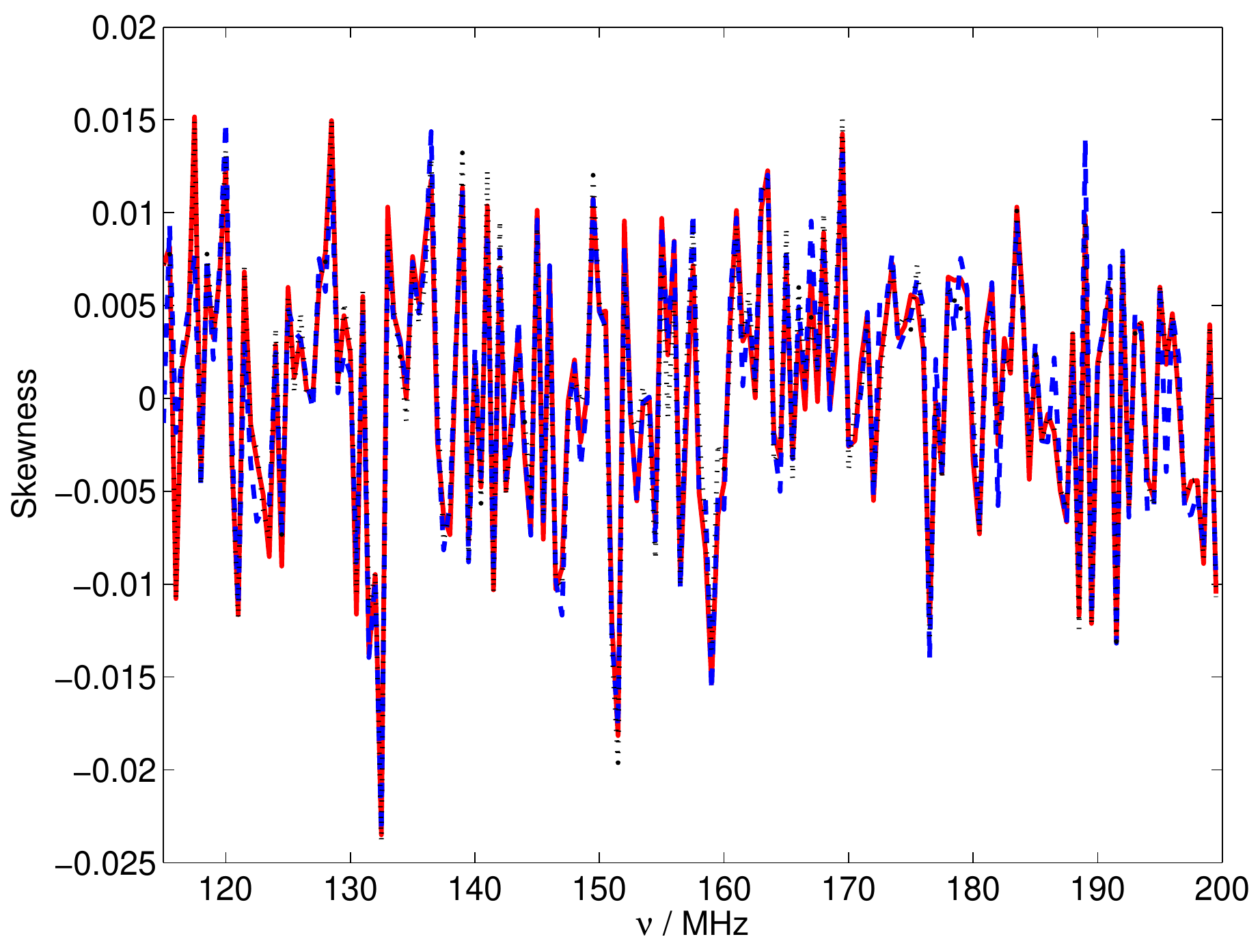}
\caption{The skewness of the simulated 21-cm signal plus noise (red), noise alone (black;dot) and residual maps (blue; dash).}
\label{skew_rest_nocs}
\end{figure}

\begin{figure}
\includegraphics[width=84mm]{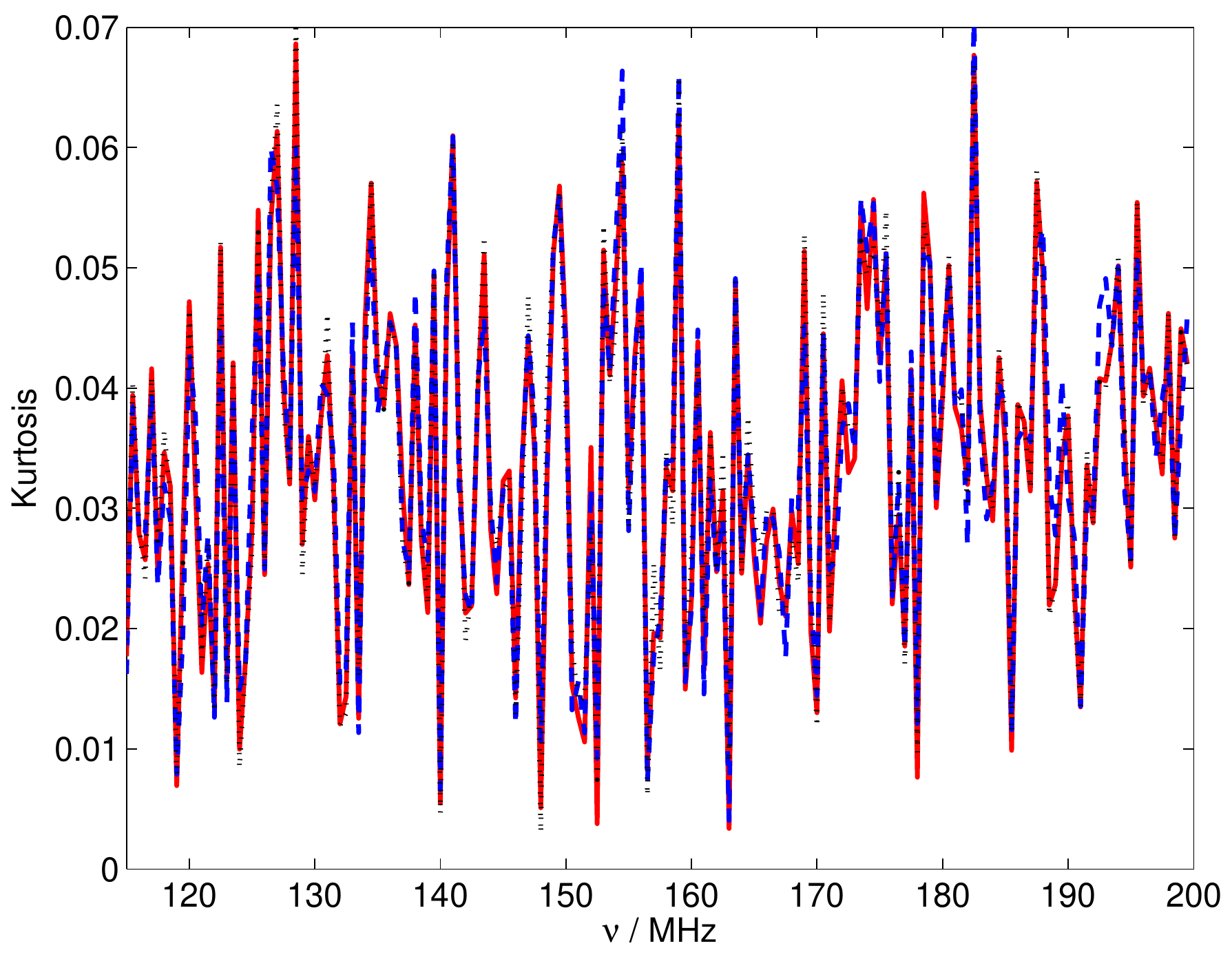}
\caption{The kurtosis of the simulated 21-cm signal plus noise (red), noise alone (black;dot) and residual maps (blue; dash).}
\label{kurt_rest_nocs}
\end{figure}

\citet{harker09a} employed a Wiener filter on the dirty residual data to denoise the images, recovering the general trends of the 21-cm skew. Kurtosis recovery proved more elusive. We present the skewness and kurtosis of the residuals cubes, Fig. \ref{skew_rest_nocs} and Fig. \ref{kurt_rest_nocs}.

The skewness and kurtosis in the residual images is recovered very well, accurately matching the simulated noise plus 21-cm signal skewness and kurtosis across the frequency range. 

\begin{figure}
\includegraphics[width=84mm]{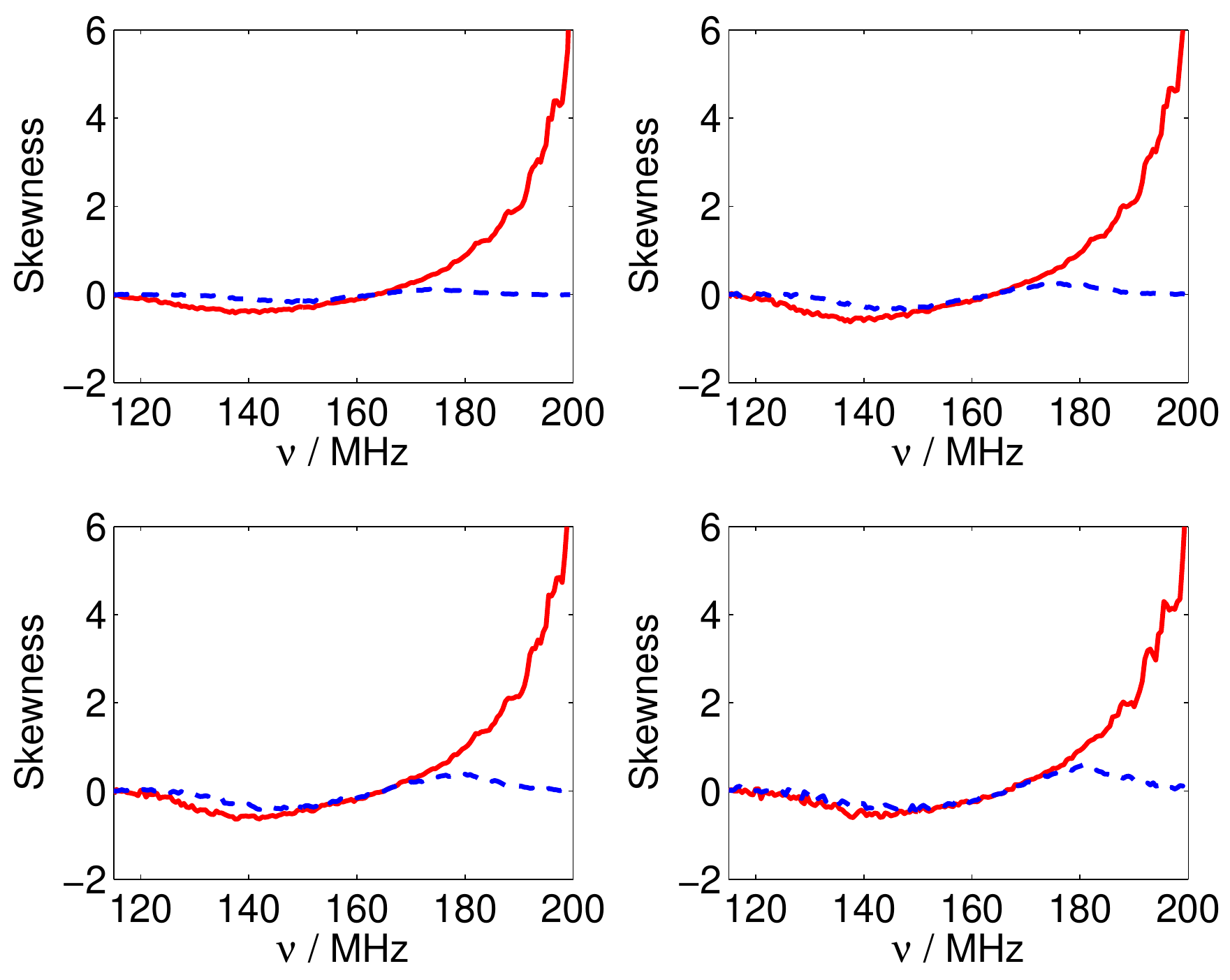}
\caption{The skewness of the simulated 21-cm signal (red; solid) and reconstructed 21-cm maps (blue; dash) for the fiducial signal and for different levels of Fourier filtering: 2,3 and 5 PSF scales (in reading order).}
\label{skew_csrec_csCHEAT}
\end{figure}

\begin{figure}
\includegraphics[width=84mm]{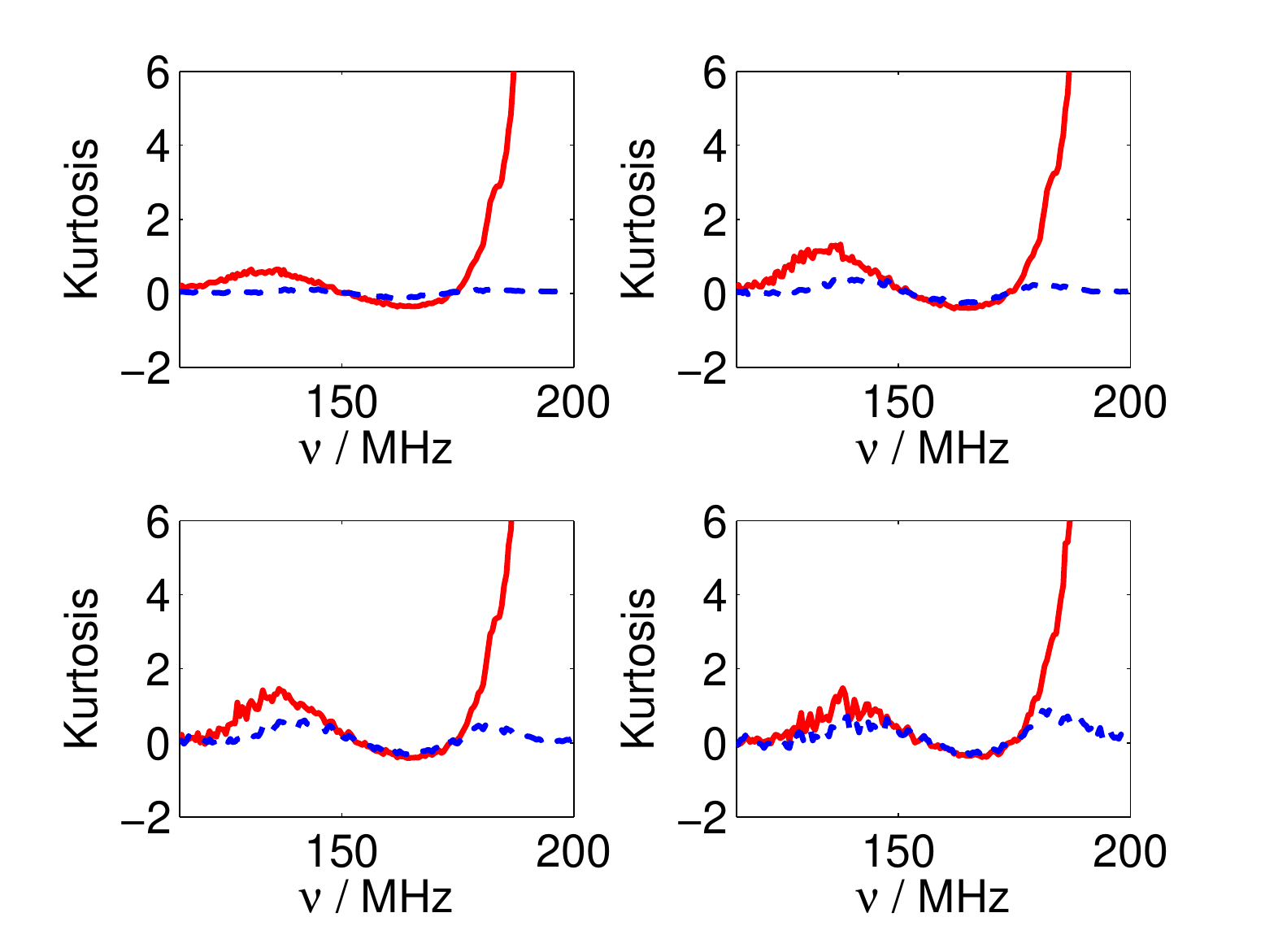}
\caption{The kurtosis of the simulated 21-cm signal (red; solid) and reconstructed 21-cm maps (blue; dash) for the fiducial signal and for different levels of Fourier filtering: 2,3 and 5 PSF scales (in reading order).}
\label{kurt_csrec_csCHEAT}
\end{figure}

We now manually subtract the noise cube from the residuals cube and plot the kurtosis/skewness of this reconstructed 21-cm signal, Fig. \ref{skew_csrec_csCHEAT} and Fig. \ref{kurt_csrec_csCHEAT}. This amounts to assuming that we know the noise distribution perfectly which, though not viable for real data, allows us an insight into the recovered signal. 

We see that the skewness dip at low frequencies is only convincingly recovered with a high level of Fourier filtering. At the high frequencies, where the cosmological signal is very small, the skewness is not recovered. The dip in kurtosis at frequency 165 MHz is somewhat recovered for Fourier filtering below 2 PSF scales while it takes up to 5 PSF scales of $k$ space filtering before the peak centred around 140MHz is recovered. For both statistics there is a divergence above frequencies of 180 MHz, where the cosmological signal is very small.

All of the results presented in this section are for \textsc{fastica} being implemented in real space. While an implementation was carried out in Fourier space, the general conclusions for all results remained the same. Though there were small local variations in, for example, the recovered power spectrum points or kurtosis values, the graphs were for all intents and purposes duplications of the real space versions and are therefore not reproduced here.

\section{Sensitivity of {\sc fastica}}
\label{limit}
So far in this paper we have assumed that the full field of view and frequency bandwidth of the simulation is input to \textsc{fastica} but we must also consider whether this method will be as successful under more constrained observations.
\subsection{Bandwidth of Observation}
\begin{figure}
\includegraphics[width=84mm]{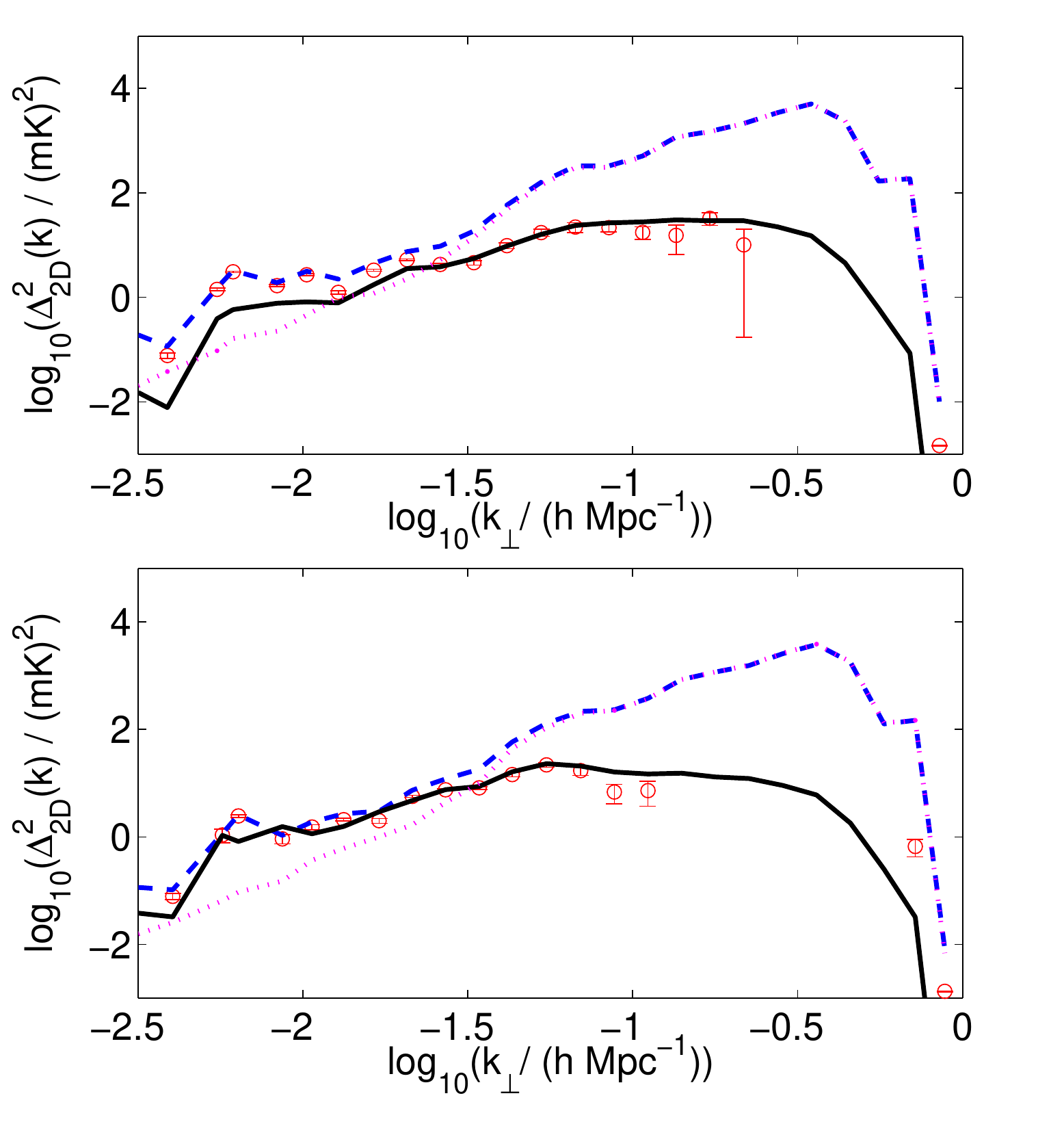}
\caption{2D power spectrum of the simulated 21-cm signal (black;solid), reconstructed 21-cm signal (red,points), residuals (blue,dash) and noise (pink,dotted) at 151 MHz (top) and 171 MHz (bottom). The graphs are for data cubes of bandwidth 42.5 MHz and ranges 115MHz - 157MHz and 157.5MHz - 199.5MHz respectively. Any error bars extending to below the x axis in linear space are shown extending to negative infinity in log space.}
\label{lbw}
\end{figure}

Firstly, we assess the sensitivity to bandwidth and split the dirty data cube into two smaller cubes of bandwidth 42.5 MHz, one from 115MHz - 157MHz and one from 157.5MHz - 199.5MHz. We perform \textsc{fastica} on each of these separately and measure the 2D power spectrum as described previously, Fig. \ref{lbw}.

We can see that even for slices at the end of the cube frequency range (i.e. Fig. \ref{lbw}, top shows a slice 6 MHz from the end of that cube) the 21-cm reconstruction is successful. The general degradation is not unexpected as the more data a separation technique has to fix the foregrounds, the better the reconstruction will be. We conclude that the method is not sensitive to the point of endangering the signal recovery, but larger bandwidths are preferable.

\subsection{Noise}
Despite the encouraging results so far, the evidence of noise leakage in the recovered maps (Fig. \ref{csrec_150}) and the variance recovery (Fig. \ref{var_fid}) motivates us to consider the sensitivity of the 21-cm statistical recovery when there is increased noise in the observation. 

\begin{figure}
\includegraphics[width=84mm]{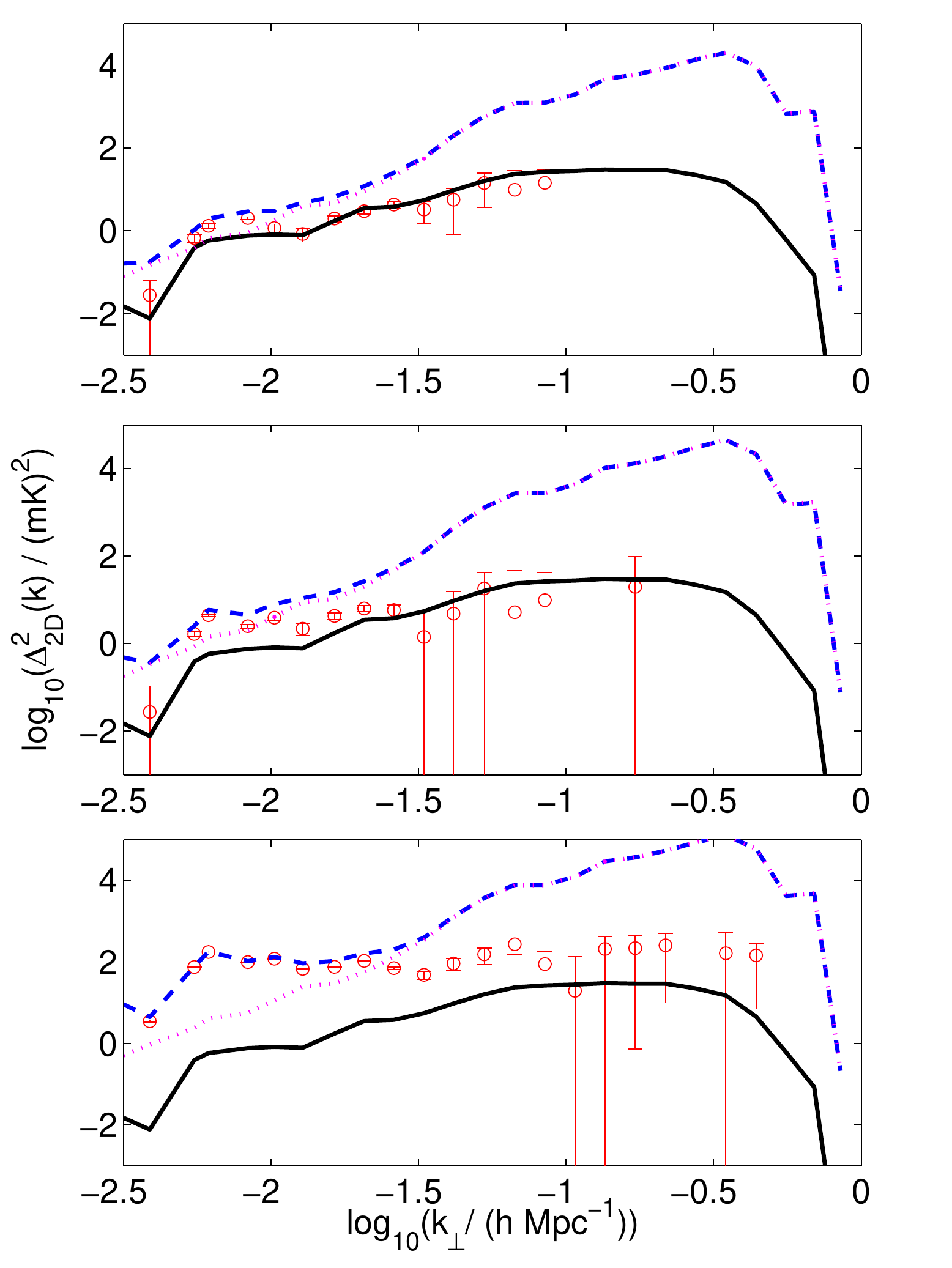}
\caption{2D power spectrum of the simulated 21-cm signal (black,solid), reconstructed 21-cm signal (red,points), residuals (blue,dash) and noise (pink,dotted) at 151 MHz, or $z$=7.30. From top to bottom, the noise simulation is set at twice, three times and five times the expected LOFAR noise respectively. Any error bars extending to below the x axis in linear space are shown extending to negative infinity in log space.}
\label{nopo}
\end{figure}

We have seen that by much reducing the expected LOFAR noise to expected SKA levels, the 21-cm  cross correlations power spectrum recovery is extremely accurate, Fig. \ref{cc_ska}. For completeness, here we set up some `worst case' scenarios, whereby we measure the recovered power spectra in the presence of two times, three times and five times the expected LOFAR noise, Fig. \ref{nopo}.

As expected, the more noise present, the less accurately the 21-cm power spectrum is recovered. For twice the expected level of noise we see the larger scales beginning to be overestimated, the extent of which worsens for three times the expected noise. For five times the expected amount of noise the power spectrum is significantly overestimated across the scale range. However we must stress that the fact that the 21-cm power spectrum is recovered across a wide scale range, even in the presence of twice the noise levels expected, can only be seen as extremely promising.

\subsection{Field of View}
\begin{figure}
\includegraphics[width=84mm]{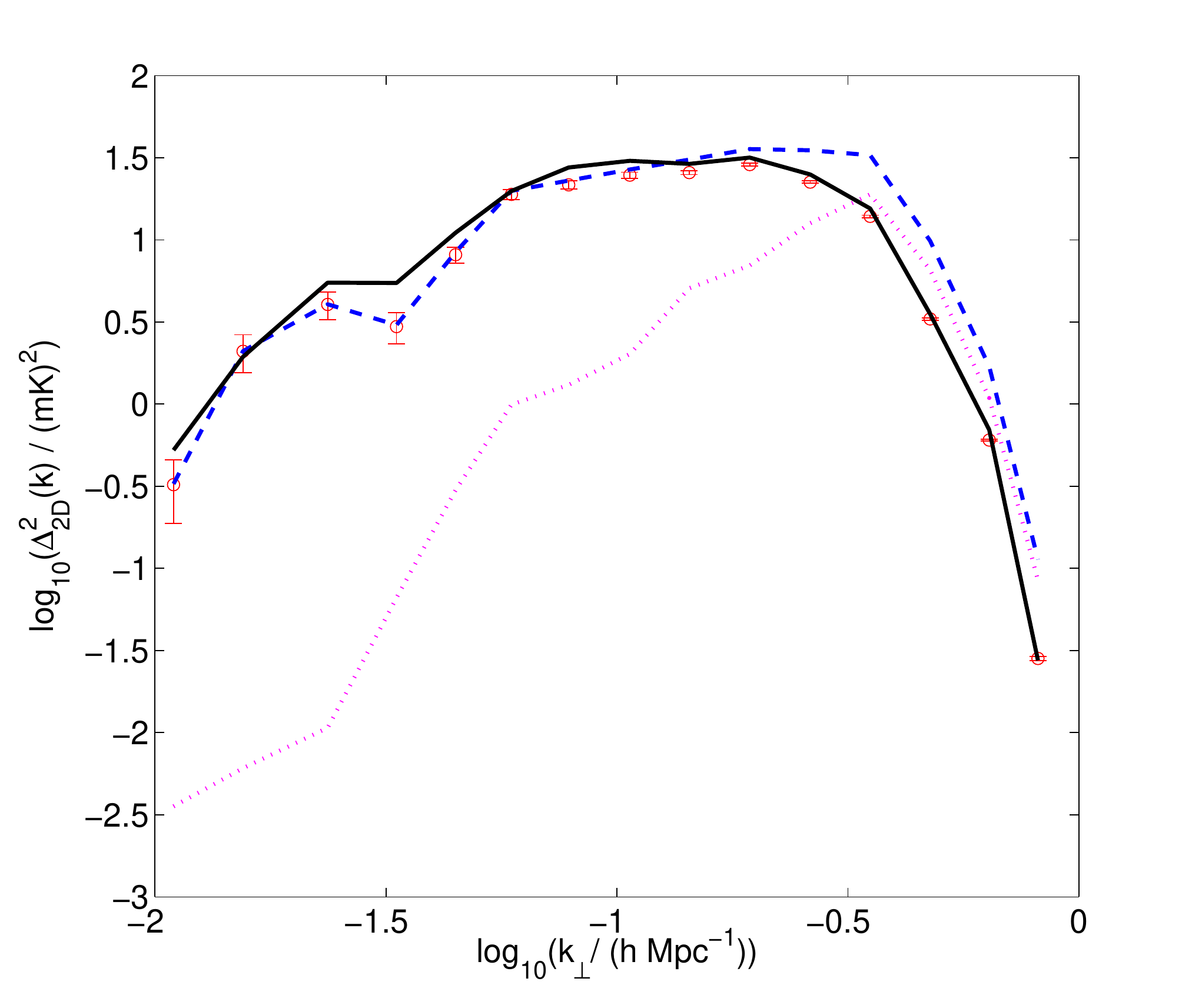}
\caption{2D power spectrum of the simulated 21-cm signal (black,solid), reconstructed 21-cm signal (red,points), residuals (blue,dash) and noise (pink,dotted) at 151 MHz, or $z$=7.30, for a 2.5$^{\circ} \times$ 2.5$^{\circ}$ field of view. Any error bars extending to below the x axis in linear space are shown extending to negative infinity in log space.}
\label{fov}
\end{figure}

In this paper we have assumed a 10$^{\circ} \times$ 10$^{\circ}$ field of view, which is at the upper limit of what we can expect for LOFAR observations. To explore the sensitivity of the analysis to the field of view, we now process a 2.5$^{\circ} \times$ 2.5 $^{\circ}$ data cube. If we had kept the noise and the resolution the same, analysing such a data cube would be plagued with noise as we would have reduced the number of pixels that we are analysing. Hence, we can choose to analyse a smaller patch in the sky with a higher resolution and same noise or decrease the noise and have similar resolution in order to have similar constraining power as the fiducial analysis and establish the effect of the sky area coverage. In actual observations a decrease in field of view and an increase in resolution would be related to the size of the stations and distribution of the stations respectively. If we had changed the resolution we would no longer correspond strictly speaking to a LOFAR case scenario. We therefore decide to decrease the field of view by a factor of 4 and enhance the signal to noise by a factor of 16. We see that the residuals are actually lower than the original 21-cm signal at the larger scales, Fig. \ref{fov}, however the 21-cm power spectrum is still well recovered at the smaller scales. We interpret this as evidence that the 21-cm signal has been mixed into the other signals by \textsc{fastica}, potentially because \textsc{fastica} did not have as many lines of sight to remove the foregrounds, making the reconstruction less accurate though still successful.

\section{Conclusions}
\label{conclusions}
We have presented a new implementation of a non-parametric foreground cleaning method using the \textsc{fastica} algorithm. \textsc{fastica} is an ICA technique which uses negentropy as a measure of non-Gaussianity. By maximising the non-Gaussianity of a signal mixture the ICs of the foregrounds can be separated. \textsc{fastica} can then reconstruct the foregrounds, with any data not considered to be part of the foregrounds forming the residuals. The residuals consist of the 21-cm signal, system noise and fitting errors.

The success of using the \textsc{fastica} method to obtain an EoR signature was tested by attempting extraction of the two main statistical markers of the EoR, the 21-cm power spectrum and variance. The rms foreground fitting error is bounded below $10$ mK across almost all of the frequency range when pixels with disproportionate errors due to unusually small foreground values are discarded. 

Once the variance of the noise has been subtracted from the variance of the residuals, an excess variance is recovered. To accurately recover the 21-cm variance it was necessary to Fourier filter the data up to about 5 times the PSF scale. In this case the excess variance accurately recovers the order and shape of the simulated 21-cm variance across the majority of the frequency range, failing only where the signal to noise is extremely low.

The 21-cm angular power spectrum and 3D power spectrum are recovered very well across a wide frequency range. 

Performing the ICA in Fourier space provides no particular advantages or disadvantages according to the statistical tests carried out in this paper. This is in contrast to other methods which have shown preference towards processing in Fourier space \citep{harker09b}. 

The \textsc{fastica} method has proved not to be robust in the presence of large amounts of noise. Though impressive results are obtained at large scales even for twice the expected levels of noise, levels above this endanger the recovery.

We have shown that \textsc{fastica} can be a competitive foreground removal technique for EoR data, though for a full treatment of the LOFAR-EoR data, the polarisation of the simulated data and a more accurate frequency dependent PSF model needs to be considered in future work. 

\section{Acknowledgments}
FBA acknowledges the support of the Royal Society via a University Research Fellowship. GH is a member of the LUNAR consortium, which is funded by the NASA Lunar Science Institute (via Cooperative Agreement NNA09DB30A) to investigate concepts for astrophysical observatories on the Moon. The authors would like to acknowledge Jacques Delabrouille and Mario Santos for useful discussion.

\bibliography{fastICA_4.0}
\bibliographystyle{mn2e}  

\end{document}